\documentclass[a4paper,11pt]{article}

\usepackage{jheppub} 

\usepackage[T1]{fontenc} 
\usepackage{tikz}
\usepackage{amsfonts}
\usepackage{physics}
\usepackage{float}
\usepackage{mathtools}
\usepackage{amsmath}
\usepackage{amsthm}
\usepackage{amssymb}
\usepackage{esint}
\usepackage{pifont}
\usepackage[numbers,sort&compress]{natbib}
\usepackage{color}
\usepackage{amsmath}
\usepackage{subfig}

\DeclareMathOperator{\arccosh}{arccosh}

\DeclareMathOperator{\arctanh}{arctanh}

\title{\boldmath Defect Extremal Surface for Reflected Entropy}

\author[a]{Tianyi Li,}
\author[a]{Ma-Ke Yuan,}
\author[a,b]{Yang Zhou}

\affiliation[a]{Department of Physics and Center for Field Theory and Particle Physics, Fudan University, Shanghai 200433, China}
\affiliation[b]{Peng Huanwu Center for Fundamental Theory, Hefei, Anhui 230026, China}

\abstract{Defect extremal surface is defined by extremizing the Ryu-Takayanagi formula corrected by the quantum defect theory. This is interesting when the AdS bulk contains a defect brane (or string). We introduce a defect extremal surface formula for reflected entropy, which is a mixed state generalization of entanglement entropy measure. Based on a decomposition procedure of an AdS bulk with a brane, we demonstrate the equivalence between defect extremal surface formula and island formula for reflected entropy in AdS$_3$/BCFT$_2$. We also compute the evolution of reflected entropy in evaporating black hole model and find that defect extremal surface formula agrees with island formula.}

\begin{document} 
\maketitle
\flushbottom
\section{Introduction}
Entanglement entropy plays an important role in recent understanding of black hole information paradox. In particular the island formula for the radiation gives Page curve \cite{Page:1993wv,Page:2013dx,Hawking:1976ra,Penington:2019npb,Almheiri:2019psf,Almheiri:2019hni,Almheiri:2019qdq,Penington:2019kki} and therefore maintains unitarity. The development relies on the quantum extremal surface formula for the fine grained entropy, which was inspired from the quantum corrected Ryu-Takayanagi formula in computing holographic entanglement entropy \cite{RT:RT-formula,HRT:HRT-formula,Lewkowycz:2013nqa,Faulkner:2013ana,Engelhardt:2014gca,Rozali:2019day,Almheiri:2019yqk,Chen:2019iro,Chen:2020wiq,Gautason:2020tmk,Anegawa:2020ezn,Hashimoto:2020cas,Hartman:2020swn,Hollowood:2020cou,Alishahiha:2020qza,Zhao:2019nxk,Chen:2019uhq,Almheiri:2019psy,Bak:2020enw,Bousso:2020kmy,Chen:2020jvn,Chen:2020tes,Hartman:2020khs,VanRaamsdonk:2020tlr,Liu:2020jsv,Langhoff:2020jqa,Balasubramanian:2020xqf,Chen:2020uac,Chen:2020hmv,Ling:2020laa,Bhattacharya:2020uun,Marolf:2020rpm,Harlow:2020bee,Nomura:2020ska,Hernandez:2020nem,Chen:2020ojn,Kirklin:2020zic,Goto:2020wnk,Hsin:2020mfa,Akal:2020twv,Numasawa:2020sty,Basak:2020aaa,Geng:2020fxl,Choudhury:2020hil,Karananas:2020fwx,Bousso:2021sji,Patrascu:2021fyg,May:2021zyu,Kawabata:2021hac,Anderson:2021vof,Bhattacharya:2021jrn,Kim:2021gzd,Hollowood:2021nlo,Miyata:2021ncm,Ghosh:2021axl,Uhlemann:2021nhu,Neuenfeld:2021bsb,Geng:2021wcq,Bachas:2021fqo,Wang:2021woy,Fallows:2021sge,Qi:2021sxb,Ahn:2021chg}. While most of the recent studies concern the entanglement entropy, which is a unique measure characterizing entanglement between two subsystems $A$ and $B$ for a pure state $\psi_{AB}$, in this paper we want to study the correlation between two subsystems in a mixed state. There are several motivations to do so: First, the entire state of black hole and radiation is not always pure. Second, understanding the rich correlations between subsystems of radiation is probably the key to understand the nature of island.

Recently {\it reflected entropy} was proposed to quantify an amount of total correlation, including quantum entanglement, for bipartite mixed states $\rho_{AB}$ based on canonical purification \cite{Dutta:2019gen,Chu:2019etd,Bao:2019zqc,Moosa:2020vcs,Bueno:2020vnx,Berthiere:2020ihq,Bueno:2020fle}. It has been shown that the holographic dual of reflected entropy is entanglement wedge cross section in AdS/CFT correspondence. Island formula for reflected entropy has been proposed in \cite{Chandrasekaran:2020qtn,Li:2020ceg} for a gravitational system coupled to a quantum bath, conjectured to be the correct one incorporating quantum gravity effects. It is surprising that a semi-classical formula can capture some key property of quantum gravity. It is therefore interesting to ask how we can justify island formula by other means. 

In this paper we propose defect extremal surface formula for reflected entropy in holographic models with defects. Defect extremal surface is defined by extremizing the Ryu-Takayanagi formula corrected by the quantum defect theory \cite{Deng:2020ent,Chu:2021gdb}. This is interesting when the AdS bulk contains a defect brane (or string). The defect extremal surface formula for reflected entropy is a mixed state generalization of that for entanglement entropy. Based on a decomposition procedure of an AdS bulk with a brane, we demonstrate in this paper the equivalence between defect extremal surface formula and island formula for reflected entropy in AdS$_3$/BCFT$_2$. We also compute the evolution of reflected entropy in evaporating black hole model and find that defect extremal surface formula agrees with island formula.

\section{Review of the model}
In this section we briefly review the holographic model we are considering. AdS/BCFT model was proposed by Takayanagi \cite{Takayanagi:2011zk}, based on the earlier work by Karch and Randall \cite{Karch:2000ct}. We slightly modify this setting by adding conformal matter on the End-of-the-World (EOW) brane in the AdS bulk. We then review the defect extremal surface (DES) which computes the entanglement entropy from the bulk description. We also review the decomposition procedure to obtain the effective boundary description by combining Randall-Sundrum reduction and Maldacena duality \cite{Deng:2020ent}, and show that the entropy computed by boundary island formula agrees with that calculated by DES formula precisely.
\subsection{AdS$_3$/BCFT$_2$}
\begin{figure}[htbp]
  \centering
  \includegraphics[scale=0.7]{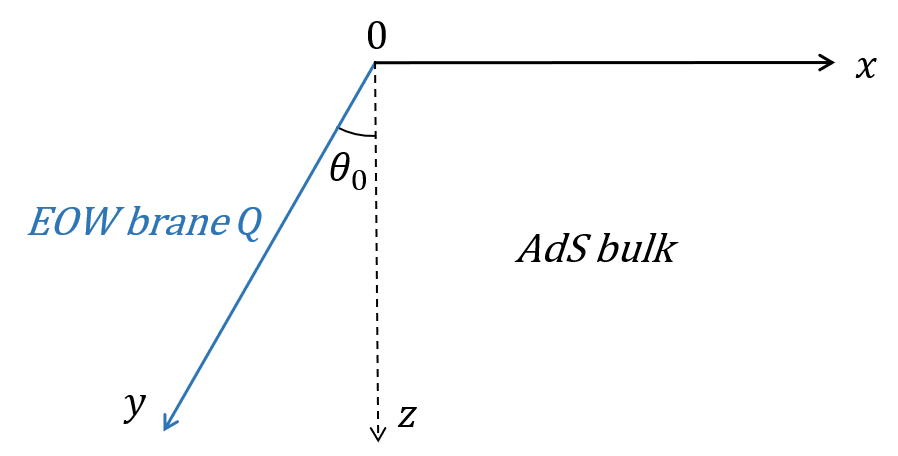}
  \caption{A fixed time slice of the AdS$_3$/BCFT$_2$ setting. The AdS$_3$ bulk has an End-of-the-World brane $Q$ which can be viewed as a defect. The holographic dual of the bulk is a BCFT defined on half space ($x>0$).}
  \label{XXX1}
\end{figure}

It was proposed that the holographic dual for $2d$ BCFT is given by a AdS$_3$ geometry with a boundary codimension-1 brane $Q$ \cite{Takayanagi:2011zk}, on which the Neumann boundary condition is imposed. The bulk action is given by
\begin{equation}
I= \frac{1}{16\pi G_N} \int_N \sqrt{-g}(R-2\Lambda)+\frac{1}{8\pi G_N} \int_Q \sqrt{-h}(K-T)\ ,
\end{equation}
where $N$ denotes the bulk and $Q$ the brane, the tension on which is denoted by a constant $T$. The variation of this action gives the Neumann boundary condition
\begin{equation}
\label{eq:nbc}
K_{a b}=(K-T) h_{a b}\ ,
\end{equation}
where $h_{a b}$ is the induced metric and $K_{a b}$ is the extrinsic curvature of the $Q$ brane. As shown in Fig.\ref{XXX1}, there are two useful coordinate systems in this geometry: $(t,x,z)$ and $(t,\rho,y)$. Their relation is given by
\begin{equation}\label{rhoy}
z=-y / \cosh \frac{\rho}{l}\ , \quad x=y \tanh \frac{\rho}{l}\ .
\end{equation}
The metric of AdS$_3$ geometry can be written as
\begin{equation}\begin{split}\label{eq1}
ds^{2}&=d\rho^{2}+l^2\cosh ^{2} \frac{\rho}{l} \cdot \frac{-dt^{2}+dy^{2}}{y^{2}}\\
&=\frac{l^2}{z^2}(-dt^2+dz^2+dx^2)\ ,
\end{split}\end{equation}
where $l$ is the AdS radius. One can also introduce the polar coordinate $\theta$ which is related with $\rho$ through $(\cos \theta)^{-1}=\cosh (\rho/l)$. Thus, $y$ can be understood as the radial coordinate. 

Suppose that the $Q$ brane is located at $\rho=\rho_0$, where $\rho_0$ is a positive constant, it is easy to show that
\begin{equation}
\label{eq:ec}
K_{a b}=\frac{\tanh \frac{\rho_0}{l}}{l} h_{a b}\ .
\end{equation}
Combining \eqref{eq:nbc} with \eqref{eq:ec}, one can find the relation between the tension $T$ and the brane location $\rho_0$,
\begin{equation}
T=\frac{\tanh \frac{\rho_0}{l}}{l}\ .
\end{equation}
Also note that since the EOW brane is located at a fixed $\rho=\rho_0$ slice, the metric induced on the brane can be easily read off from (\ref{eq1}), which is a AdS$_2$ geometry
\begin{equation}
ds_{brane}^2 = l^2\cosh ^{2} \frac{\rho_0}{l} \cdot \frac{-dt^{2}+dy^{2}}{y^{2}}\ .
\end{equation}
\subsection{Bulk defect extremal surface}\label{sec_des}
\begin{figure}[htbp]
  \centering
  \includegraphics[scale=0.7]{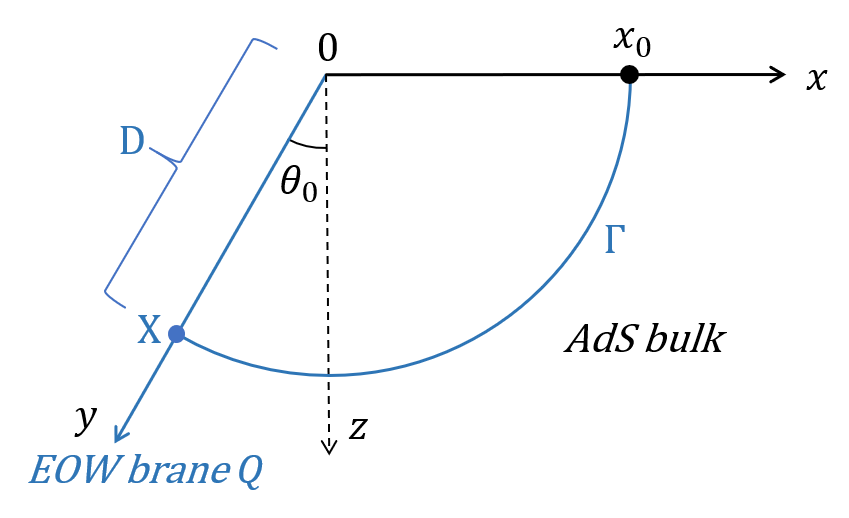}\\
  \caption{The RT surface $\Gamma$ for an interval $I := [0; x_0]$ that contains the boundary. When the boundary of the BCFT is contained in the interval, $\Gamma$ will extend to the EOW brane, on which conformal matter is distributed.}
  \label{XXX2}
\end{figure}

In \cite{Takayanagi:2011zk} it was proposed that the entanglement entropy of an interval $[0,x_0]$ on the BCFT can be calculated holographically by the bulk RT surface, as illustrated in Fig.\ref{XXX2}. Now if we take the brane as a defect in the bulk and add conformal matter on it, it is obvious that one should include the contribution from these matter when calculating the entanglement entropy. Thus the RT formula should be modified to the defect extremal surface (DES) formula, which is given by \cite{Deng:2020ent}
\begin{equation}\label{DES}
S_{\mathrm{DES}}=\min _{\Gamma, X}\left\{\operatorname{ext}_{\Gamma, X}\left[\frac{\operatorname{Area}(\Gamma)}{4 G_{N}}+S_{\text {defect }}[D]\right]\right\}, \quad X=\Gamma \cap D\ ,
\end{equation}
where $\Gamma$ is a co-dimension two surface in AdS bulk and $X$ is the lower dimensional entangling surface determined by the intersection of $\Gamma$ and the defect $D$. 
The entanglement entropy for an interval $I:=[0,x_0]$ calculated by DES formula (\ref{DES}) is given by
\begin{equation}\label{bgen}
S_{\text{DES}}=\frac{c}{6} \log \frac{2x_0}{\epsilon}+\frac{c}{6} \arctanh \sin \theta_0 + \frac{c'}{6} \log {\frac{2l}{\epsilon_y \cos\theta_0}}\ ,
\end{equation}
where $c'$ is the central charge of the brane conformal field theory. It is worth noting that the defect contribution on the AdS$_2$ brane is a constant, $S_{\text {defect}}=\frac{c'}{6} \log {\frac{2l}{\epsilon_y \cos\theta_0}}$, therefore the defect extremal surface is the same as the RT surface shown in Fig.\ref{XXX2}.

\subsection{Boundary island formula}
The boundary island formula proposed in \cite{Almheiri:2019hni} can be understood as the boundary dual of the above bulk DES formula. In order to see it, one has to first derive an effective $2d$ description from the AdS$_3$ bulk with an End-of-the-World brane. This process can be done through the combination of {\it partial Randall-Sundrum reduction} and AdS/CFT correspondence, which we briefly review as follows. For the detail discussion, see \cite{Deng:2020ent}.

\par
\begin{figure}[htbp]
	\centering
		\subfloat[The bulk description]{
			\begin{minipage}{7cm}
			\centering
			\includegraphics[scale=0.45]{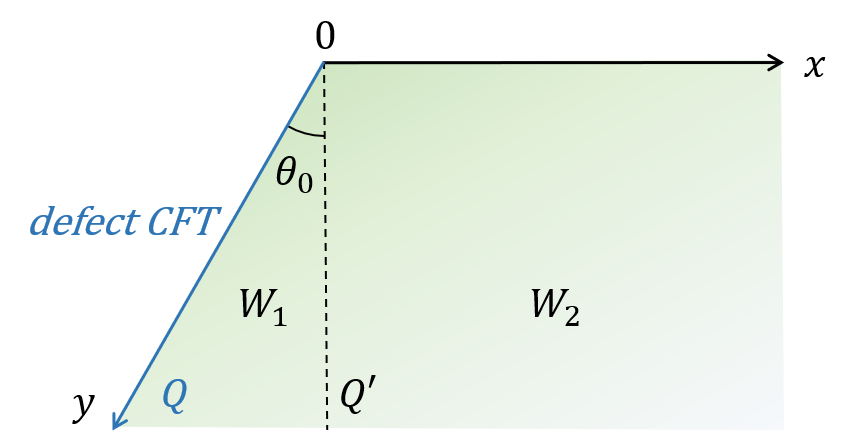}
			\end{minipage}
		}
		\subfloat[The effective boundary description]{
			\begin{minipage}{7cm}
			\centering
			\includegraphics[scale=0.45]{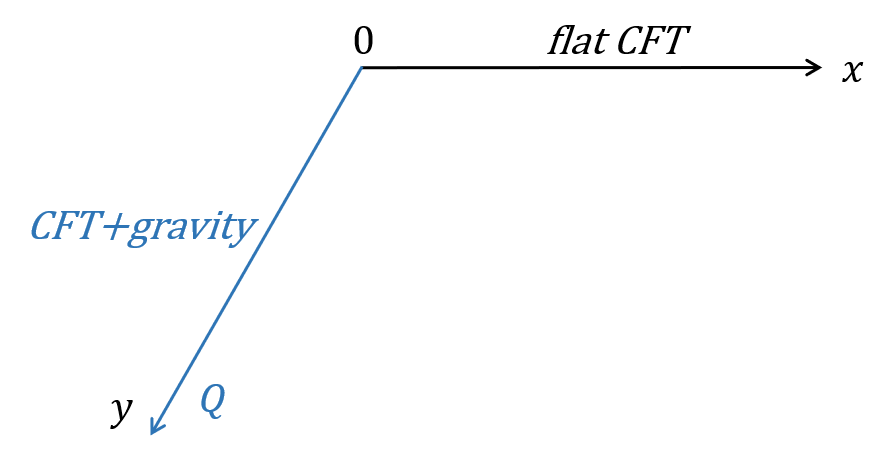}
			\end{minipage}
		}
	\caption{(a) The bulk geometry is an AdS$_3$ bulk with an EOW brane on which there is a defect CFT. The imaginary boundary $Q'$ (dotted line) divides the bulk into two parts, $W_1$ and $W_2$. (b) The boundary description consists of a flat CFT gluing to a CFT coupled to gravity. This $2d$ effective theory is obtained by applying partial Randall-Sundrum reduction and AdS/CFT correspondence to $W_1$ and $W_2$ respectively.}
	\label{XXX3}
\end{figure}

 Let us first decompose the AdS$_3$ bulk into two parts $W_1$ and $W_2$ by inserting an imaginary codimension-1 boundary $Q'$ orthogonal to the asymptotic boundary (Fig.\ref{XXX3}(a)). There are no physical degrees of freedom living on $Q'$, on which transparent boundary condition is imposed. The $2d$ description of $W_2$ can be obtained by simply replacing itself with the BCFT with zero boundary entropy due to AdS/CFT correspondence. To find the $2d$ description of $W_1$, one can take a partial Randall-Sundrum reduction along $\rho$ direction, which leads to a gravity theory on the EOW brane $Q$. Together with the conformal matter on $Q$, one gets a $2d$ gravity theory coupled to brane CFT. The brane CFT and the half-space flat CFT are glued together with transparent boundary condition, which is essentially the dual of the imaginary boundary $Q'$. The final effective $2d$ description is shown in Fig.\ref{XXX3}(b).

Having obtained an effective boundary description, one can use the island formula to compute entanglement entropy. For instance, the entanglement entropy of an interval $[0,L]$ in the flat CFT region with $c=c'$ is calculated by island formula as
\begin{equation}\label{114}
S^{\text{bdy}}=\text{ext}_a\big\{S_{\text{gen}}(a)\big\}=\text{ext}_a\big\{S_{\text{area}}
(y=a)+S_{\text{matter}}([-a,L])\big\}\ ,
\end{equation}
where $a$ is the boundary of the island on the brane. The area term in the case of $d=2$ is given by
\begin{equation}\label{art}
\frac{1}{4G_{N}^{(2)}}=\frac{\rho_0}{4G_{N}}=\frac{c}{6}\text{arctanh} \sin \theta_0\ ,
\end{equation}
where we have used the fact that the CFT central charge on the asymptotic boundary is related to the bulk Newton constant by $c=\frac{3l}{2G_N}$. Inserting (\ref{art}) into (\ref{114}) and extremizing over $a$, one gets the final entanglement entropy
\begin{equation}\begin{split}
S^{\text{bdy}}&=\frac{c}{6} \arctanh \sin \theta_0 + \frac{c}{6}\log \frac{4x_0l}{\cos\theta_0\epsilon\epsilon_y}\\
&=\frac{c}{6} \log \frac{2x_0}{\epsilon}+\frac{c}{6} \arctanh \sin \theta_0 + \frac{c}{6} \log \frac{2l}{\epsilon_y \cos\theta_0}\ .
\end{split}\end{equation}
One can see that the island result is exactly the same as the bulk DES result supposing $c=c'$. In summary, defect extremal surface gives a holographic derivation of the island formula for entanglement entropy. 

Our main goal in this paper is to generalize the defect extremal surface to compute reflected entropy, which is the mixed state analogy of entanglement entropy. We will also check the consistency with the boundary island formula of reflected entropy conjectured in earlier works \cite{Chandrasekaran:2020qtn,Li:2020ceg}. We will start by working in a static model and then move to the time dependent case, reflecting the physics of black hole evaporation.
\section{The island formula for reflected entropy and its bulk dual}
\begin{figure}[htbp]
	\centering
	\includegraphics[scale=0.45]{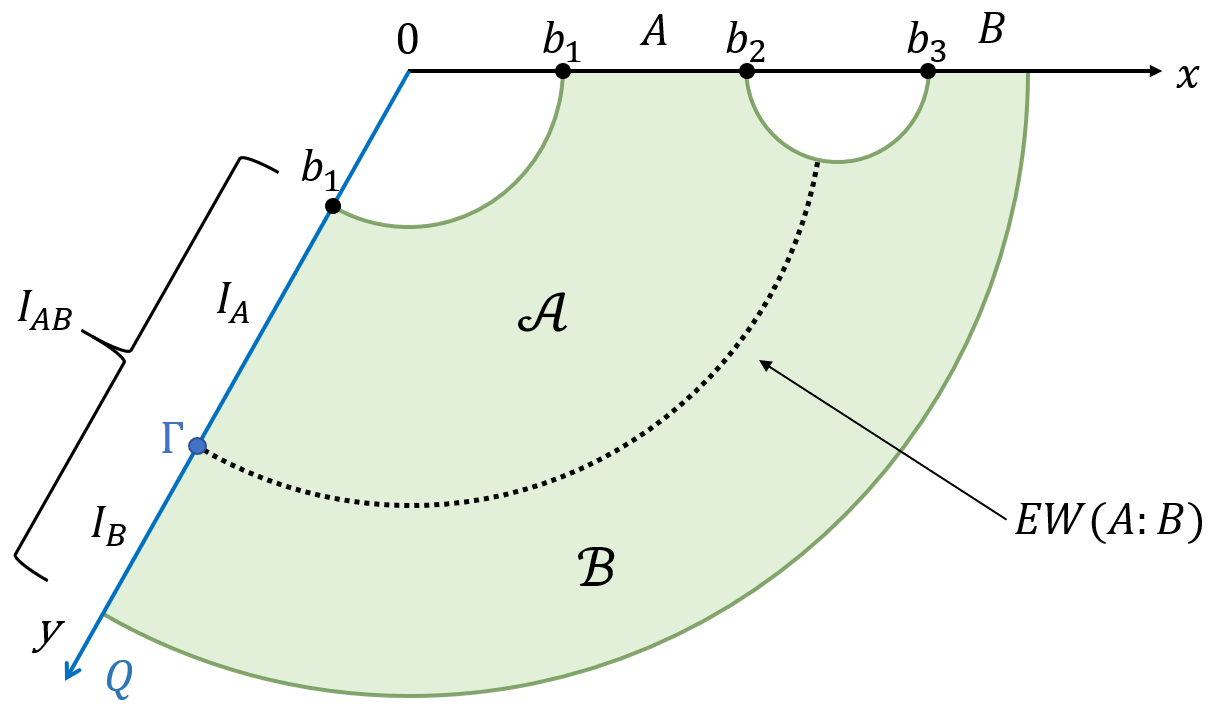}
	\caption{An illustration of the island formula for reflected entropy. The entanglement wedge of $A\cup B$ is the green shaded region, where there exists a minimal entanglement wedge cross section $EW(A:B)$ that splits the bulk into $\mathcal{A}$ and $\mathcal{B}$, which corresponds to the boundary intervals $A$ and $B$ respectively. The intersection of $EW(A:B)$ and $Q$ is denoted by $\Gamma$, which is the island cross section from boundary point of view.} 
	\label{XXX4}
\end{figure}
The island formula for entanglement entropy can be generalized to calculate reflected entropy as follows. As shown in Fig.\ref{XXX4}, for two intervals $A$ and $B$ in asymptotic boundary, the reflected entropy between them is given by \cite{Chandrasekaran:2020qtn,Li:2020ceg}
\begin{equation}\label{SRbdy}
S_R^{\text{bdy}}(A:B)=\text{min }\text{Ext}_{\Gamma}\left\{S_R^{(\text{eff})}(A\cup I_A:B\cup I_B)+\frac{\text{Area}[\Gamma]}{2G_N}\right\}\ ,
\end{equation}
where $I_A$ and $I_B$ together make up the entanglement island $I_{AB}$ and its \emph{island cross section} $\partial I_A\cap \partial I_B$ is denoted by $\Gamma$. Note that when computing the reflected entropy between $A$ and $B$, one first needs to find $I_{AB}$, the entanglement island of the total system. Then the location of the island cross section is determined by minimizing the functional (\ref{SRbdy}) over $\Gamma$.
\par In the spirit of the duality discussed above, the reflected entropy could also be computed in the bulk description as seen in Fig.\ref{XXX4}, which is given by
\begin{equation}\label{SRbulk}
S_R^{\text{bulk}}(\mathcal{A}:\mathcal{B})=\text{min }\text{Ext}_{\text{EW}}\left\{S_R^{(\text{eff})}(\mathcal{A}:\mathcal{B})+\frac{\text{Area}[\text{EW}(A:B)]}{2G_N}\right\}\ ,
\end{equation}
where the \emph{entanglement wedge cross section}, denoted by $\text{EW}(A:B)$, splits the entanglement wedge of $A\cup B$ into two parts $\mathcal{A}$ and $\mathcal{B}$ in the bulk. Since the conformal matter is only located on the End-of-the-World brane, the first term in (\ref{SRbulk}) boils down to the effective reflected entropy between $I_A$ and $I_B$ on the brane. We call this formula the defect extremal cross section formula for reflected entropy. In the following section, we would compute the reflected entropy of two intervals in a static time slice of the AdS$_3$/BCFT$_2$ model using the two formulas above and check the consistency of the results.

\section{Reflected entropy in a static time slice}
In this section, we compute the reflected entropy between the intervals $A=[b_1,b_2]$ and $B=[b_3,\infty]$ in a fixed time slice of the AdS$_3$/BCFT$_2$ model. There are three possible phases for the computation.
\subsection{The first phase}
\begin{figure}[htbp]
	\centering 
		\subfloat[The first phase]{
			\begin{minipage}{7cm}
			\centering
			\includegraphics[scale=0.4]{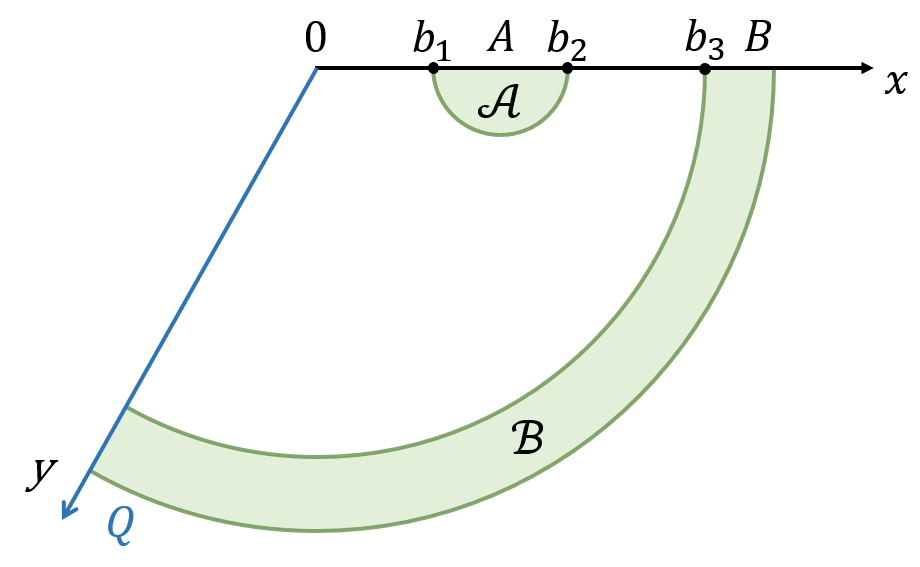}
			\end{minipage}
		}
		\subfloat[The second phase]{
			\begin{minipage}{7cm}
			\centering
			\includegraphics[scale=0.4]{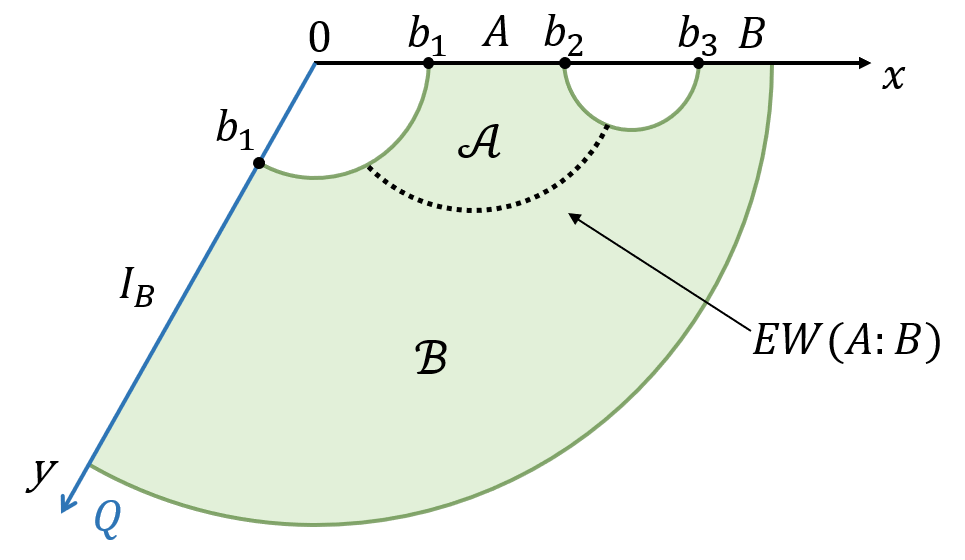}
			\end{minipage}
		}
	\caption{(a) In the first phase, $\mathcal{A}$ and $\mathcal{B}$ are naturally separated. There is no entanglement wedge cross section. (b) In the second phase, $EW(A:B)$ does not touch the $Q$ brane and thus no island cross section appears in the dual boundary description.}
	\label{XXX5}
\end{figure}

This is the most trivial phase, which corresponds to disconnected entanglement wedges. As shown in Fig.\ref{XXX5}(a), the entanglement wedge of $A\cup B$ is naturally divided into two parts $\mathcal{A}$ and $\mathcal{B}$. From the bulk description, there is no entanglement wedge cross section, so the area term in (\ref{SRbulk}) vanishes. The effective entropy term also vanishes since conformal matter is only located on the $Q$ brane. Thus, the final reflected entropy in this phase is simply zero,
\begin{equation}
S_R^{\text{bulk}}(\mathcal{A}:\mathcal{B})=S_R^{\text{bdy}}(A:B)=0\ .
\end{equation}

\subsection{The second phase}
In this phase the entanglement wedge cross section $EW(A:B)$ hangs on the extremal surfaces\footnote{As shown in \ref{sec_des}, defect extremal surfaces always have the same shape as RT surfaces.} of $A\cup B$, as shown in Fig.\ref{XXX5}(b). Apparently there is no island cross section on the brane, so the reflected entropy in the boundary description reduces to $S_R^{\text{(eff)}}(A:B\cup I_B)$. This calculation is similar to eq.(4.40) in \cite{Dutta:2019gen}, which gives
\begin{equation}
S_R^{\text{bdy}}(A:B)=\frac{2c}{3}\log{(\frac{1+\sqrt{1-x}}{\sqrt{x}})}\ , \quad x=\frac{(b_2-b_1)(b_3+b_1)}{(b_3-b_1)(b_1+b_2)}\ .
\end{equation}
In the bulk description, the reflected entropy between $\mathcal{A}$ and $\mathcal{B}$ is simply the area of $EW(A:B)$ divided by $2G_N$. Using eq.(20) in \cite{Takayanagi:2017knl}, one can show that
\begin{equation}
S_R^{\text{bulk}}(\mathcal{A}:\mathcal{B})=\frac{c}{3}\log{(1+2z+2\sqrt{z(z+1)})}\ , \quad z = \frac{2b_1(b_3-b_2)}{(b_2-b_1)(b_3+b_1)}\ .
\end{equation}
It is easy to find exact agreement between the two results above.
\subsection{The third phase}\label{the_third_phase}
As illustrated in Fig.\ref{XXX4}, This phase corresponds to a non-trivial island cross section on the brane, which divides the entanglement island into $I_A$ and $I_B$. Following the same method used in \cite{Chandrasekaran:2020qtn}, one can obtain the reflected entropy from the boundary point of view, which is given by
\begin{small}
\begin{equation}\label{SR2}
S_R^{\text{bdy}}(A:B)=\frac{c}{3}\left[\log{(b_3+a)}+\log{(b_2+a)}-\log{(b_3-b_2)}+\log{\frac{2l}{a\epsilon_y\cos\theta_0}}+\arctanh \sin\theta_0 \right]\ ,
\end{equation}
\end{small}
where $a$ is the coordinate of the island cross section and $\epsilon_y$ is the UV regulator of the brane. Extremizing (\ref{SR2}) over $a$ leads to the solution
\begin{equation}
a_0=\sqrt{b_2 b_3}\ .
\end{equation}
Now we would like to compute the reflected entropy from the bulk side. First we need to compute the effective reflected entropy of the BCFT on the brane, i.e. $S_R^{(\text{eff})}(I_A:I_B)$. To do this, one must evaluate the correlation functions of twist operators $\sigma_i$
\begin{equation}\label{SRBCFT}
S_R^{(\text{eff})}(I_A:I_B)=\lim\limits_{m\to1}\lim\limits_{n\to1}\frac{1}{1-n}\log{\frac{\prod_i \Omega_i^{2h_i} \left< \sigma_{g_A}(b_1) \sigma_{g_A^{-1}g_B}(a) \right>_{BCFT^{\bigotimes mn}}}{\Omega(b_1)^{2 h_m n}\left< \sigma_{g_m}(b_1)\right>^n_{BCFT^{\bigotimes m}}}}\ ,
\end{equation}
where $\Omega_i$ is the associated conformal factor and $h_i$ is the conformal dimension of $\sigma_i$. This formula comes from a double replica trick where $mn$ copies of a single system are glued together. The operators like $\sigma_{g_A}$ and $\sigma_{g_B}$ are twist operators inserted at the boundary of associated intervals, which impose specific permutations of these copies. For instance, in order to compute the reflected entropy between two disjoint intervals $A$ and $B$, one has to insert $\sigma_{g_A}$, $\sigma_{g^{-1}_A}$, $\sigma_{g_B}$ and $\sigma_{g^{-1}_B}$ at the four endpoints of the intervals. Another type of useful operator is a composite operator like $\sigma_{g_A^{-1}g_B}$, which is the dominant operator exchanged between $\sigma_{g_A^{-1}}$ and $\sigma_{g_B}$. For more detailed discussion, we refer to \cite{Dutta:2019gen}. The conformal dimensions are given by\footnote{The conformal dimensions of $g_A$, $g^{-1}_A$, $g_B$ and $g^{-1}_B$ are equal. We may use notations like $h_A$ or $h_{AB}$ for convenience in following sections.}
\begin{equation}
	h_{g_A}=\frac{cn}{24}(m-\frac{1}{m})\ , \quad h_{g_A^{-1}g_B}=\frac{c}{12}(n-\frac{1}{n})\ , \quad h_m=\frac{c}{24}(m-\frac{1}{m})\ .
\end{equation}
One can read off the conformal factor from the induced metric on the brane, i.e. $ds^2_{brane}=\Omega^{-2}(y)ds^2_{flat}$,
\begin{equation}\label{conformal_factor}
\Omega(y)=|\frac{y\cos\theta_0}{l}|\ .
\end{equation}
The conformal invariance fixes the form of one-point function on a flat BCFT, which is given by \cite{Sully:2020pza}
\begin{equation}\label{1point}
\left<  \sigma_{g_m}(b_1) \right>_{BCFT^{\bigotimes m}}=\frac{1}{|2b_1/ \epsilon_y|^{2h_m}}\ .
\end{equation}
We have set the boundary entropy $\log{g_b}=0$ here and in the following computation. The correlator in (\ref{SRBCFT}) has two possible channels: the OPE and the boundary operator expansion (BOE). 

\begin{figure}[htbp]
	\centering
		\subfloat[]{
			\begin{minipage}{7cm}
			\centering
			\includegraphics[scale=0.6]{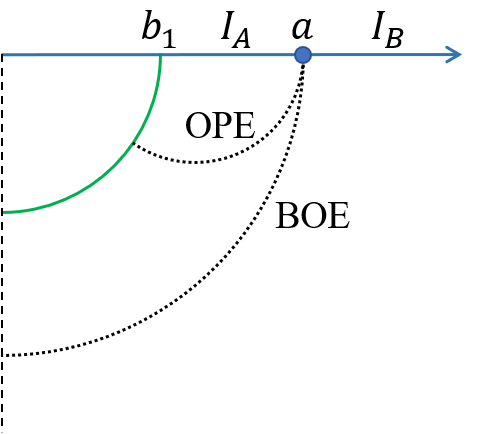}
			\end{minipage}
		}
		\subfloat[]{
			\begin{minipage}{7cm}
			\centering
			\includegraphics[scale=0.6]{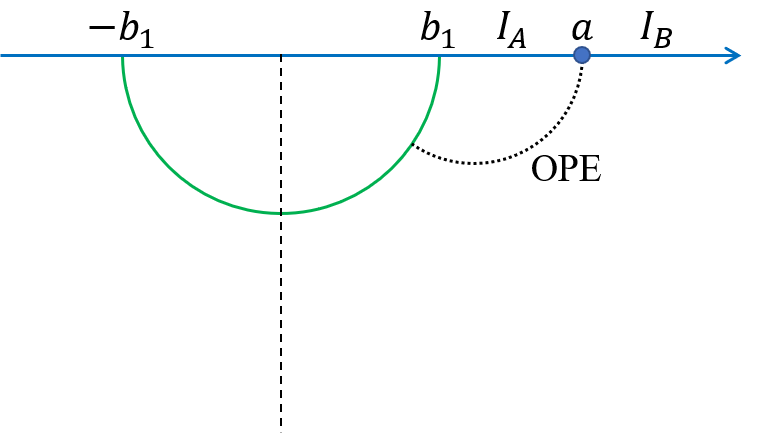}
			\end{minipage}
		}
	\caption{(a) Holographic illustration of possible channels. The brane BCFT has a holographic bulk dual cut off by an End-of-the-World brane represented by the dashed vertical line. The RT surface for $I_A\cup I_B$ is represented by the green arc. The correlators in (\ref{SRBCFT}) have two possible channels, which correspond to the area of two different wedge cross sections separating $I_A$ and $I_B$ represented by dashed arcs. From the field theory side, the BOE channel corresponds to a product of two one-point BCFT correlators. (b) The OPE channel of the two-point BCFT correlator is equivalent to a three-point function on a whole CFT.}
	\label{channels}
\end{figure}

\paragraph{BOE channel}When this channel dominates, as shown in Fig.\ref{channels}(a), one can break the two-point correlators into a product of two one-point correlators on a flat BCFT. Recalling (\ref{1point}), the result reads
\begin{equation}\label{BOE2point}
\left< \sigma_{g_A}(b_1) \sigma_{g_A^{-1}g_B}(a) \right>_{BCFT^{\bigotimes mn}}=\frac{\epsilon_y^{2(h_{g_A}+h_{g_A^{-1}g_B})}}{(2b_1)^{2h_{g_A}}(2a)^{h_{g_A^{-1}g_B}}}\ .
\end{equation}
Inserting (\ref{1point})(\ref{BOE2point}) into (\ref{SRBCFT}), one can obtain the reflected entropy in this channel
\begin{equation}\label{SR0}
S_R^{(\text{eff})}(I_A:I_B)=\frac{c}{3}\log{\frac{2l}{\epsilon_y\cos{\theta_0}}}\ .
\end{equation}
Note that this reflected entropy is twice the entanglement entropy of $I_B$, which coincides with the holographic picture in Fig.\ref{channels}(a).

\paragraph{OPE channel}In this channel, as illustrated in Fig.\ref{channels}(b), the corresponding two-point BCFT correlators is actually equivalent to a three-point function on a flat CFT
\begin{equation}\label{2point2}
\left< \sigma_{g_A}(b_1) \sigma_{g_A^{-1}g_B}(a) \right>_{BCFT^{\bigotimes mn}}= \left< \sigma_{g_B^{-1}}(-b_1) \sigma_{g_A}(b_1)  \sigma_{g_A^{-1}g_B}(a)    \right>_{CFT^{\bigotimes mn}} \ . 
\end{equation}
According to (C.9) in \cite{Dutta:2019gen}, this three-point function is given by
\begin{equation}\label{opecoef}
\left< \sigma_{g_B^{-1}}(-b_1) \sigma_{g_A}(b_1)  \sigma_{g_A^{-1}g_B}(a)    \right>_{CFT^{\bigotimes mn}}=(2m)^{-4h_n}(a^2-b_1^2)^{-4h_n}(2b_1)^{-4nh_m+4h_n} \ , 
\end{equation}
where $h_n=\frac{c}{24}(n - \frac{1}{n})$. Using (\ref{SRBCFT})(\ref{2point2})(\ref{opecoef}) one can obtain the reflected entropy in this channel
\begin{equation}\label{SRinfty}
S_R^{(\text{eff})}(I_A:I_B)=\frac{c}{3}\log\frac{2l}{\xi\epsilon_y\cos\theta_0}\ , \quad \xi=\frac{2ab_1}{a^2-b_1^2}\ . 
\end{equation}
The valid regime of (\ref{SR0}) and (\ref{SRinfty}) can be determined by a critical value $\xi_c$ at which the dominant channel switches to the other. Thus, it is easy to show that OPE channel dominates when $\xi>1$, otherwise the BOE channel dominates, assuming large central charge limit. Next we would calculate the extremal length of the geodesic $EW(A:B)$ in Fig.\ref{XXX4}. For a geodesic touching a fixed point on the brane with coordinate $a$, the length formula is given in appendix \ref{appendix1}, which we denote by $L(a)$. In summary, one can write the reflected entropy from the bulk point of view as
\begin{equation}
\label{bt2}
\begin{split}
S_R^{\text{bulk}}(\mathcal{A}:\mathcal{B})=
\begin{cases}\frac{c}{3}\log{\frac{2l}{\epsilon_y\cos{\theta_0}}}+\frac{L(a)}{2G_N}\ ,\quad &\xi<1\\
\frac{c}{3}\log{\frac{2l}{\xi\epsilon_y\cos{\theta_0}}}+\frac{L(a)}{2G_N}\ ,\quad &\xi>1\ .
\end{cases}
\end{split}
\end{equation}
The last step is to extremize (\ref{bt2}) over $a$ to find the minimal value. By extremizing $S_R(\mathcal{A}:\mathcal{B})$ with respect to $a$, we find that $\partial_a S_R>0$ when $\xi>1$, so there is no extremal solution. When $\xi<1$, $\partial_a S_R=0$ leads to the extremal solution
\begin{equation}\label{extremalsol}
a_0=\sqrt{b_2 b_3}\ .
\end{equation}
Employing (\ref{Lmin}), one gets the final result of reflected entropy in the bulk description
\begin{equation}\label{SRbulkfinal}
S_R^{\text{bulk}}(\mathcal{A}:\mathcal{B}) = \frac{c}{3}\arccosh \frac{b_3+b_2}{b_3-b_2} + \frac{c}{3}\arccosh\frac{1}{\cos{\theta_0}} + \frac{c}{3} \log{\frac{2l}{\epsilon_y\cos\theta_0}}\ .
\end{equation}
By using the relation $\arccosh x = \log(x+\sqrt{1+x^2})$, it is easy to see that (\ref{SRbulkfinal}) exactly equals (\ref{SR2}) with $a$ replaced by $\sqrt{b_2 b_3}$.
\section{Time dependent reflected entropy in black hole evaporation}
In this section we study a time dependent AdS$_3$/BCFT$_2$ setting. We will see that an eternal black hole emerges in this setting from the boundary description. We have seen that for a static time slice the reflected entropy calculated by the island formula in the boundary description agrees with that calculated by defect extremal cross section in the bulk description. In this section we will show that the agreement holds in the time dependent case too. We find that the reflected entropy between two parts of the black hole interior goes down with time and finally vanishes. The reflected entropy between left radiation and right radiation jumps at Page time when the phase transition occurs. We also find that the reflected entropy between single side radiation and black hole as a function of time has a Page curve behavior. 
\subsection{The emergence of a $2d$ eternal black hole}
\begin{figure}[htbp]
  \centering
  \includegraphics[scale=0.5]{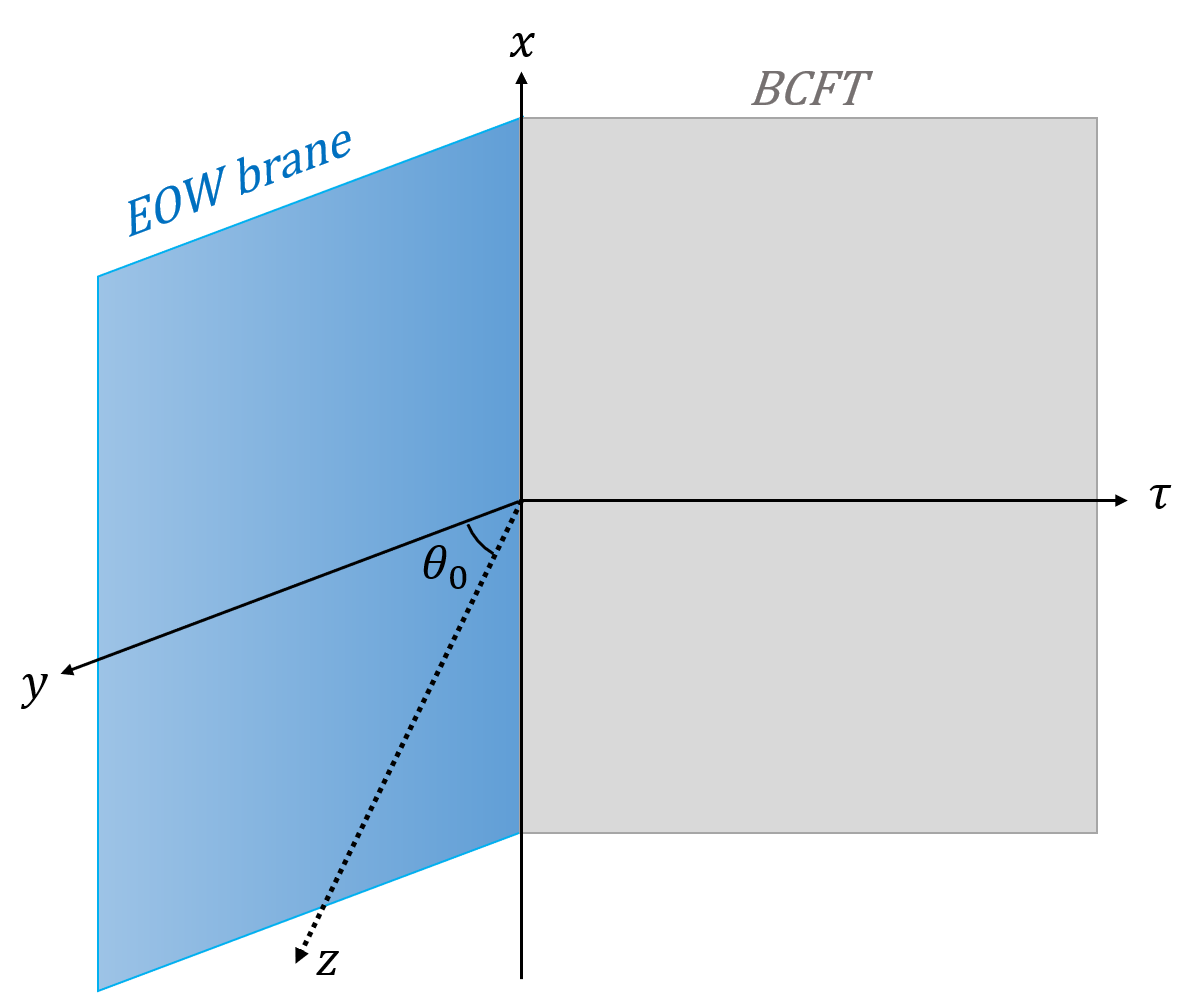}
  \caption{Holographic dual of a Euclidean BCFT defined on half spacetime ($\tau>0$). Note that $\tau$ and $x$ coordinates are exchanged, which is different from the common definition.}
  \label{XXX6}
\end{figure}

We first review how a 2-dimensional eternal black hole emerges from the AdS$_3$/BCFT$_2$ model. Recall that the holographic dual of a $2d$ BCFT can be considered as an AdS$_3$ geometry with a codimension-1 EOW brane. The metric of the Euclidean AdS$_3$ bulk is given by
\begin{equation}\label{original_coordinates}
\begin{split}
ds^{2}&=d\rho^{2}+l^2\cosh ^{2} \frac{\rho}{l} \cdot \frac{d\tau^{2}+dy^{2}}{y^{2}}\\
&=\frac{l^2}{z^2}(d\tau^2+dz^2+dx^2)\ .
\end{split}\end{equation}
As illustrated in Fig.\ref{XXX6}, the bulk geometry is bounded by the BCFT defined on a half spacetime ($\tau>0$) and an EOW brane located at $\tau=-z\tan \theta$. Note that there is no essential difference between space and time in Euclidean spacetime. Thus, the choice of $\tau$ and $x$ coordinates is just for the convenience of calculation with no physical interpretation.

\begin{figure}[htbp]
  \centering
  \includegraphics[scale=0.5]{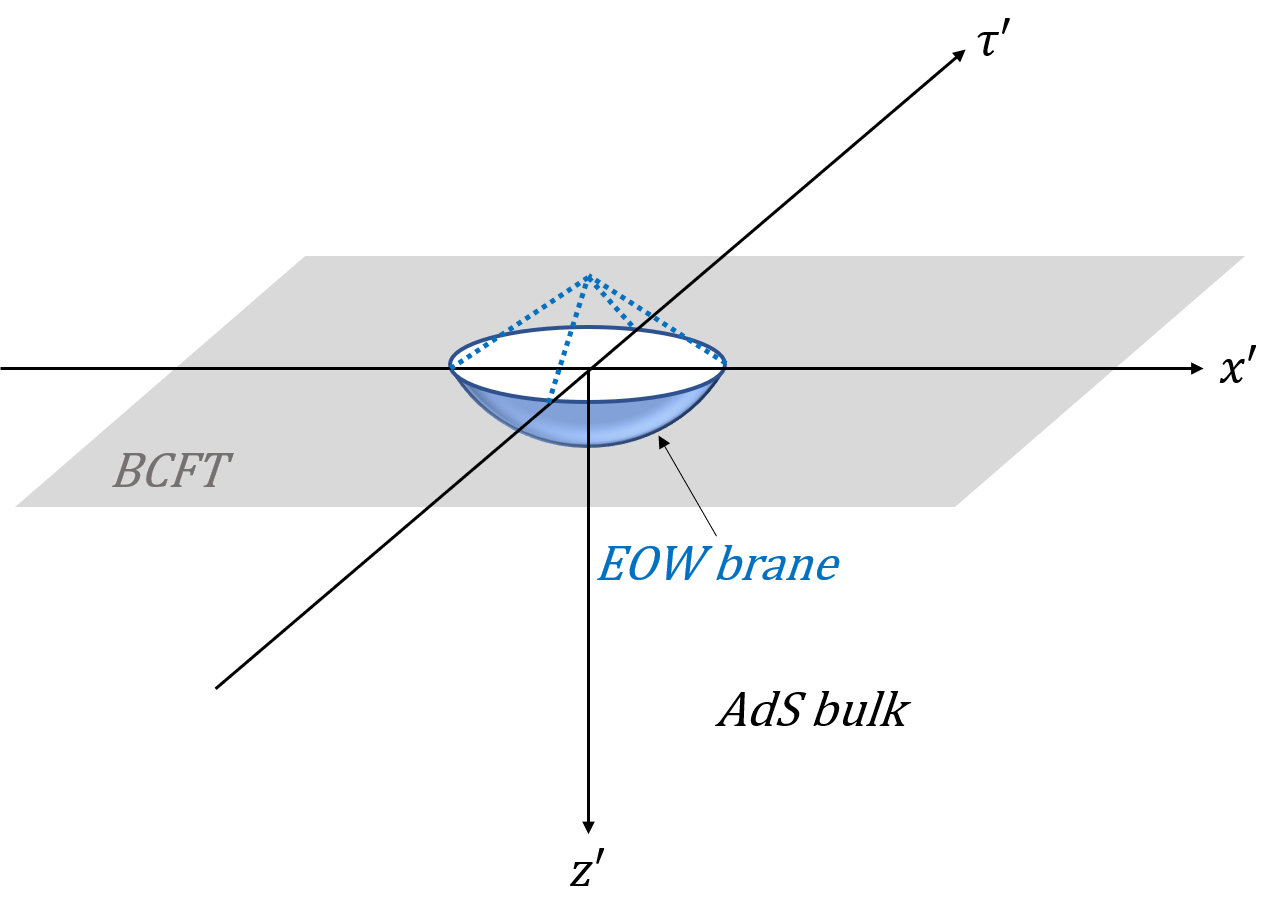}
  \caption{Holographic dual of the BCFT after the conformal transformation. The boundary of the BCFT is mapped to a circle and the EOW brane is mapped to a part of sphere.}
  \label{XXX7}
\end{figure}

In order to find a physical interpretation, one can use a set of conformal transformations
\begin{equation}
\label{sct}
\begin{split}
\tau&=\frac{2(x'^2+\tau'^2+z'^2-1)}{(\tau'+1)^2+x'^2+z'^2}\ ,\\
x&=\frac{4x'}{(\tau'+1)^2+x'^2+z'^2}\ ,\\
z&=\frac{4z'}{(\tau'+1)^2+x'^2+z'^2}\ ,
\end{split}
\end{equation}
so that the boundary of the BCFT is mapped to a circle 
\begin{equation}\label{euclidean_boundary}
x'^2+\tau'^2=1\ ,
\end{equation}
and the EOW brane is mapped to a part of sphere 
\begin{equation}
(z'+\tan\theta)^2+x'^2+\tau'^2=\sec^2\theta\ ,
\end{equation}
while the metric is preserved, as shown in Fig.\ref{XXX7}.

\begin{figure}[htbp]
  \centering
  \includegraphics[scale=0.35]{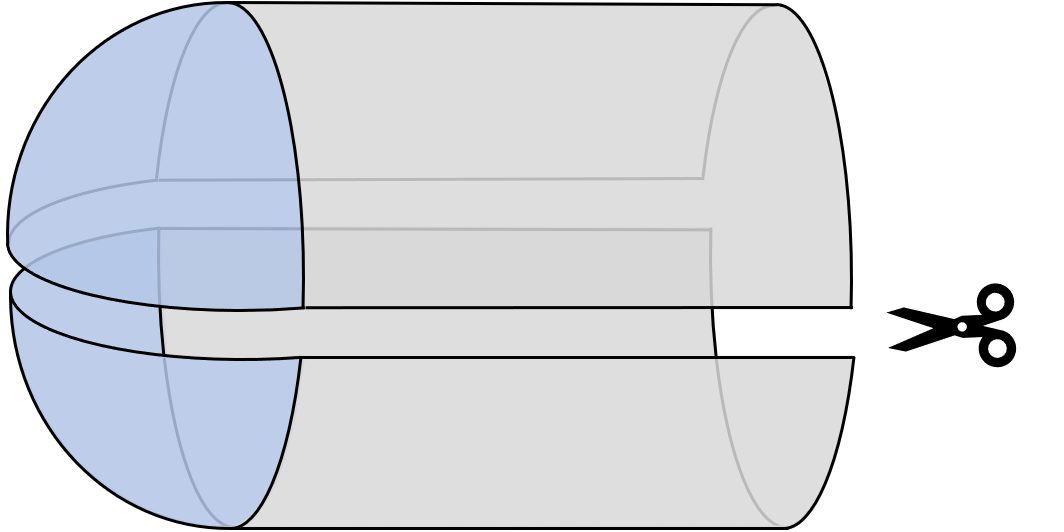}
  \caption{The initial state is prepared by cutting the path integral defined on the cylinder.}
  \label{XXX8}
\end{figure}

Using the decomposition method introduced in previous sections, one can obtain a 2-dimensional effective boundary description, which is a gravitational region on the EOW brane surrounded by a bath CFT (Fig.\ref{XXX7}). It is natural to employ radial quantization here, which means that the angle direction in polar coordinates of $x'-\tau'$ plane is taken to be a Euclidean time coordinate. One can therefore introduce another set of coordinates $(X,\phi)$ in which the bath CFT is mapped onto a cylinder shown in Fig.\ref{XXX8},
\begin{equation}\label{euclidean_prime_coordinate}
x'= e^{X}\cos\phi\ ,\quad \tau'=e^{X}\sin\phi\ .
\end{equation}

From the effective boundary description, we can imagine there is a path integral of a 1+1 dimensional CFT defined on the surface of this cylinder. Cutting off this path integral in the middle of the cylinder, one gets a state on the truncation surface prepared by the path integral on half of the cylinder. We take this slice as our initial data and then evolve it in Lorentz spacetime. 

\begin{figure}[htbp]
  \centering
  \includegraphics[scale=0.35]{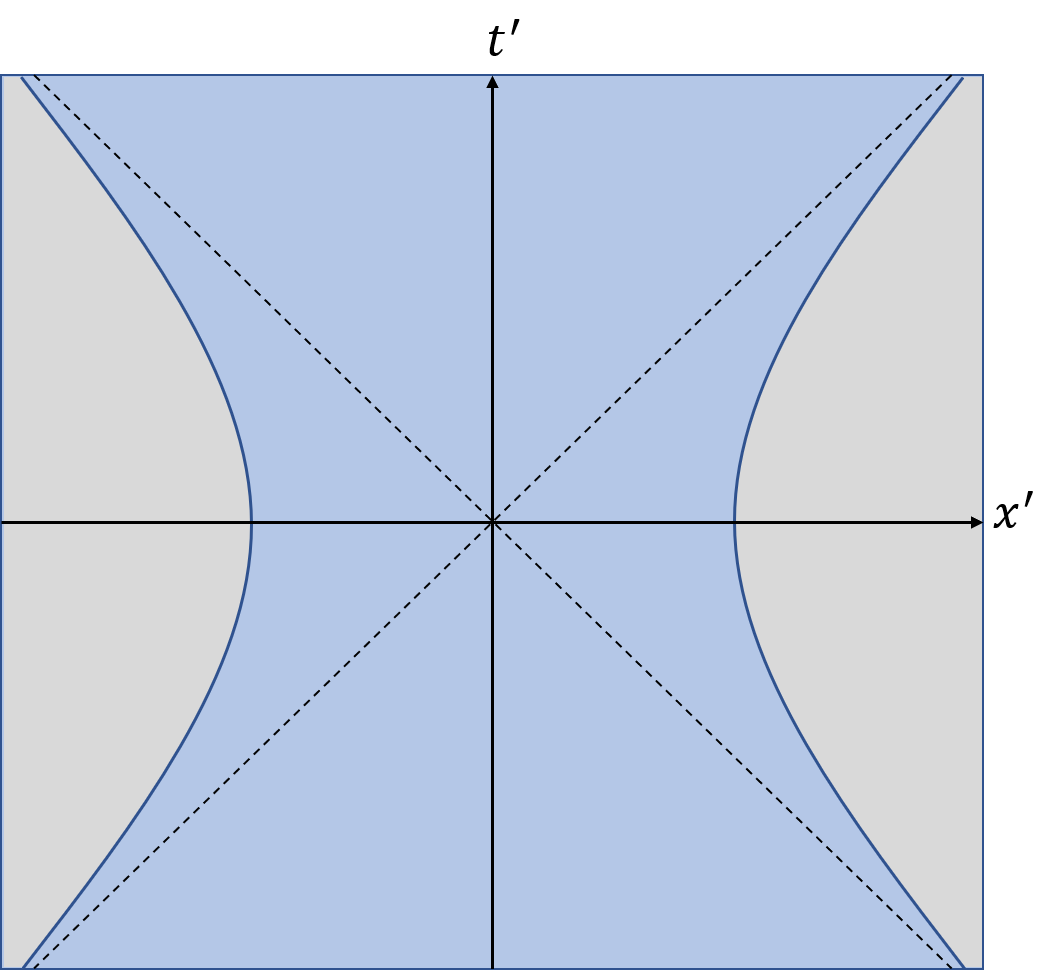}
  \caption{The Lorentz geometry viewed along $z$-axis. The boundary of the EOW brane (blue region) becomes hyperbolic curves. The horizon on the brane is represented by dashed lines.}
  \label{XXX9}
\end{figure}

The Lorentz geometry can also be understood by analytical continuing the Euclidean time $\tau'\to it'$ for the original metric, as shown in Fig.\ref{XXX9}. Note that the Euclidean boundary of the EOW brane (\ref{euclidean_boundary}) becomes $x'^2-t'^2=1$ in the Lorentz spacetime. One can then introduce a coordinate system $(X, T)$ corresponding to a Rindler observer who moves along a constant-acceleration path on the right patch of this Lorentz spacetime
\begin{equation}\label{rindler}
x'= e^{X}\cosh T\ ,\quad t'=e^{X}\sinh T\ ,
\end{equation}
then the metric in these coordinates takes the form as Rindler space, which is just the near-horizon geometry of a black hole. \footnote{The static observers ($r=\text{constant}$) in Schwarzschild are related to constant-acceleration paths ($X=\text{constant}$) in this Rindler space.} This procedure is done in the right Rindler patch, and we can do the similar thing for the left patch due to left/right $\mathbb{Z}_2$ symmetry. Thus, one finally obtains a 1+1 dimensional two-sided black hole coupled to the bath CFT which receives radiation emitted from the black hole.
\par Actually, the black hole horizon can be determined as the light-like curves on the EOW brane, the equation of which is given by
\begin{equation}
x'=\pm t'\ ,\quad z'={1-\sin\theta\over \cos\theta}\ ,
\end{equation} which asymptotes to the boundary of the EOW brane~\cite{Rozali:2019day}, $x'^2-t'^2=1$, when $t'\to \infty$. The horizon viewed along $z$-axis is shown in Fig.\ref{XXX9}.
\subsection{The reflected entropy between black hole interiors}
\subsubsection{Boundary description}
\begin{figure}[htbp]
	\centering
	\includegraphics[scale=0.5]{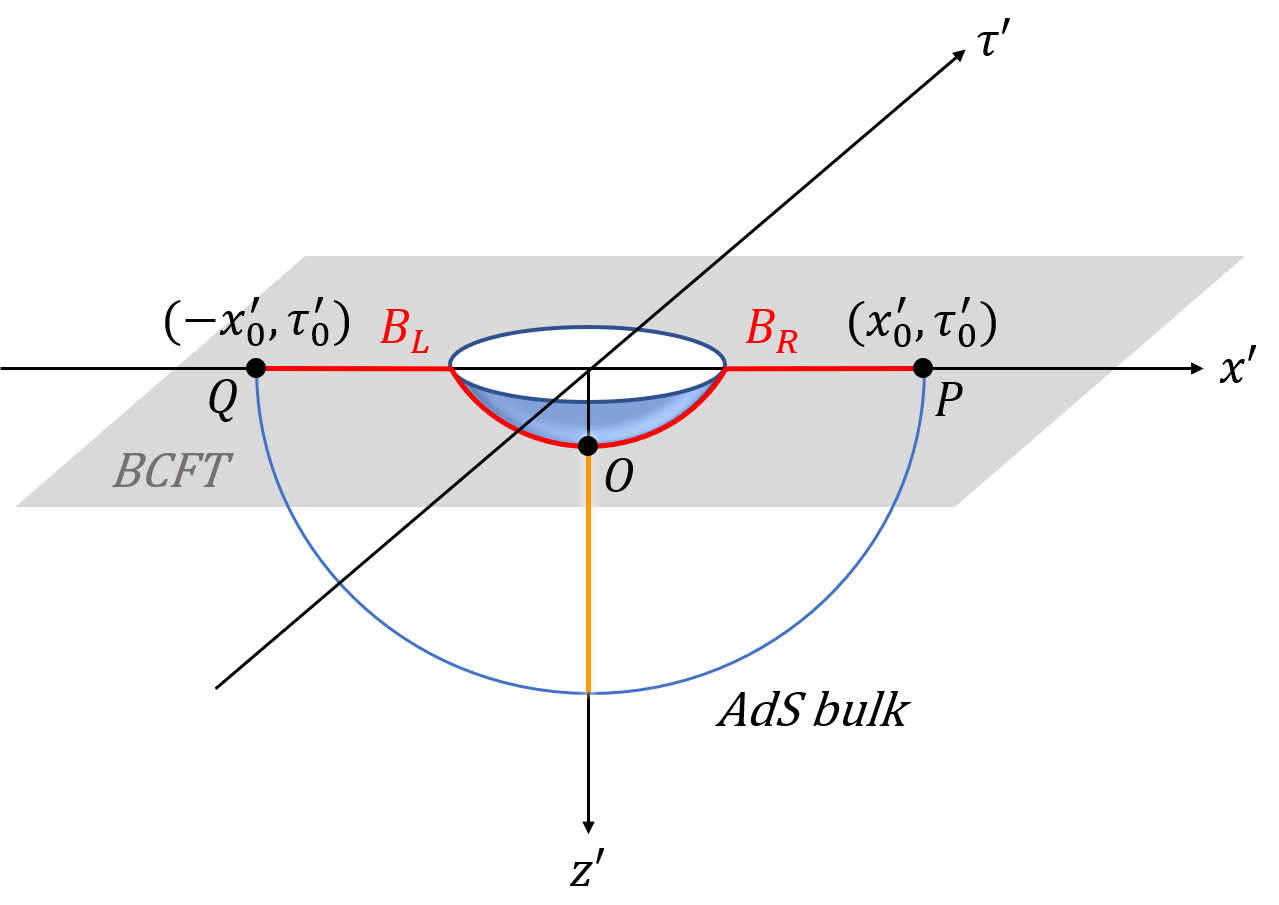}
	\caption{The connected phase of the extremal surface for the black hole. In this phase, the extremal surface (blue curve) is a geodesic connecting the endpoints of the black hole region. The cross section of the entanglement wedge is a geodesic (orange curve) connecting the extremal surface and the EOW brane, which is determined by minimizing the generalized reflected entropy (the terms in bracket in (\ref{SRbulk})) between the left and right bulk. Note that from the bulk point of view, there is conformal matter distributed on the EOW brane, which is treated as a part of the bulk.}
	\label{XXX10}
\end{figure}
Now we study the time dependent reflected entropy from the $2d$ effective boundary point of view after the decomposition procedure. We follow the setting of \cite{Chu:2021gdb} that the black hole region is defined as a space-like slice with endpoints at $(t_0', x_0')$ and $(t_0', -x_0')$. For simplicity we will do the computation in Euclidean geometry and then analytically continue the result to Lorentz spacetime. The two endpoints of the black hole interior for some constant time slice then become $(\tau_0',x_0')$ and $(\tau_0',-x_0')$. There are two phases of the extremal surface for the black hole region \cite{Chu:2021gdb}, one is connected and the other is disconnected. 
\par The connected phase shown in Fig.\ref{XXX10} corresponds to the case where there is no island contribution. Here we need to calculate the reflected entropy between the interval $|OP|$ and $|OQ|$, where $O$ is an extremal point dynamically moving on the EOW brane. Similar to the previous section, it is easy to do the computation in Euclidean geometry and then analytically continue the result to Lorentz spacetime. The endpoints in Euclidean geometry is illustrated in Fig.\ref{XXX10}. One can compute $S_R^{(\text{eff})}(|OP|:|OQ|)$ by three-point correlation functions of twist operators $\sigma_i$
\begin{equation}\label{SReffbdy_bh}
	S_R^{(\text{eff})}(|OP|:|OQ|)=\lim\limits_{m\to1}\lim\limits_{n\to1}\frac{1}{1-n}\log{\frac{ \Omega_O^{2h_{g_A^{-1}g_B}} \left< \sigma_{g_A}(P) \sigma_{g_A^{-1}g_B}(O) \sigma_{g_B^{-1}}(Q)\right>_{CFT^{\bigotimes mn}}}{ \left< \sigma_{g_m}(P)\sigma_{g_m}(Q)\right>^n_{CFT^{\bigotimes m}}}}\ ,
\end{equation}
where the conformal factor associated with the point $O$ is $\Omega_O=|\frac{y\cos\theta}{l}|$. By employing (C.9) in \cite{Dutta:2019gen}, the numerator is given by
\begin{equation}\label{numerator}
	\left< \sigma_{g_A}(P) \sigma_{g_A^{-1}g_B}(O) \sigma_{g_B^{-1}}(Q)\right>_{CFT^{\bigotimes mn}} = (2m)^{-4h_n}|OQ|^{-4h_n}|OP|^{-4h_n}|PQ|^{-4nh_m+4h_n}\ ,
\end{equation}
and the denominator reads
\begin{equation}\label{denominator}
	\left<\sigma_{g_m}(P)\sigma_{g_m}(Q)\right>^n_{CFT^{\bigotimes m}}=|PQ|^{-4nh_m}\ ,
\end{equation}
where $|OP|=|OQ|=\sqrt{(\tau_0+y)^2+x_0^2}$ and $|PQ|=2x_0$ in the $2d$ boundary description. Inserting (\ref{numerator})(\ref{denominator}) into (\ref{SReffbdy_bh}) and combining the area term (\ref{art}), one gets
\begin{equation}\label{SRbdy_bh}
	S_R^{\text{bdy}} = \frac{c}{3} \left[ \log[x_0^2+(\tau_0+y)^2] - \log x_0 + \log \frac{l}{y\cos\theta} + \arctanh \sin\theta \right].
\end{equation}
The extremal solution is given by $y=\sqrt{\tau_0^2+x_0^2}$. Thus the final reflected entropy is
\begin{small}
	\begin{equation}\label{final_result}
		S_R^{\text{bdy}} = \frac{c}{3}\left[ \log[x_0^2+(\tau_0+\sqrt{\tau_0^2+x_0^2})^2]-\log x_0 +\log \frac{l}{\sqrt{\tau_0^2+x_0^2}\cos\theta} + \arctanh \sin\theta \right].
	\end{equation}
\end{small}
Rewriting (\ref{final_result}) in Rindler coordinates $(X, T)$, and using $\arctanh \sin\theta = \log(\frac{1+\sin\theta}{1-\sin\theta})^{1/2}$, one can get the final result
\begin{small}
	\begin{equation}\label{bulk_final_1}
		S_R^{\text{bulk}}(B_L:B_R)=\frac{c}{3}\left[ \log \frac{e^{2X_0}-1+\sqrt{4e^{2X_0}\cosh^2 T+(e^{2X_0}-1)^2}}{2e^{X_0}\cosh T} + \log \frac{\cos\theta}{1-\sin\theta}+ \log\frac{2l}{\epsilon_y\cos\theta}\right],
	\end{equation}
\end{small}
where $X_0$ is a fixed constant specifying the boundary of the black hole region for a constant $T$ slice in Rindler coordinates. Note that the reflected entropy in this connected phase is a function that decreases over time.
\begin{figure}[htbp]
	\centering
	\includegraphics[scale=0.5]{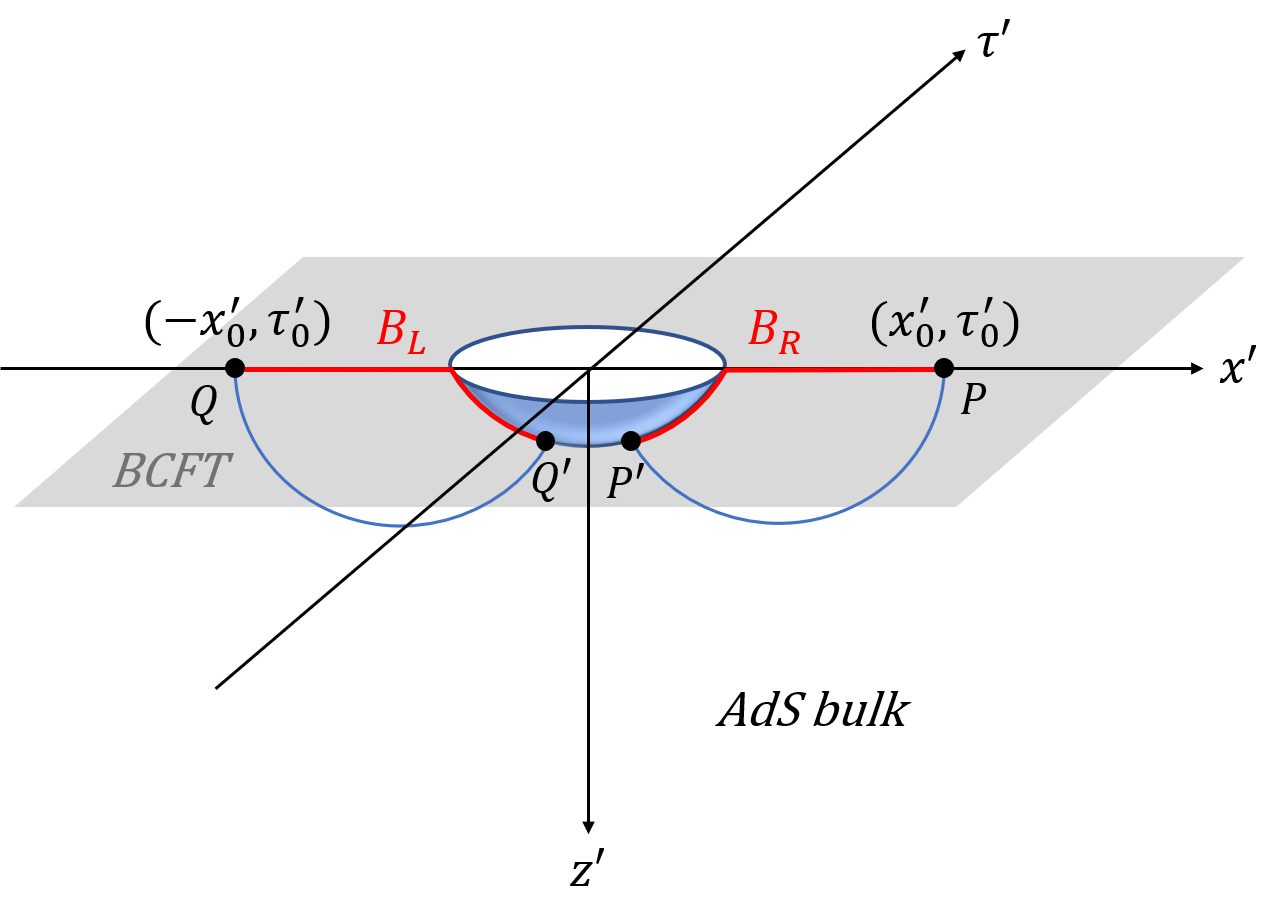}
	\caption{The disconnected phase of the extremal surface. In this phase, the extremal surfaces of the black hole region are two disconnected geodesics. The left and right parts of the black hole are naturally separated by an island appearing in the middle.}
	\label{XXX12}
\end{figure}
\par For the disconnected phase (Fig.\ref{XXX12}), there is no contribution from the area term since the two regions don't intersect. Therefore the reflected entropy is simply the effective term between $|QQ'|$ and $|PP'|$, which can be calculated from four-point correlators
\begin{equation}\label{514}
	S_R^{\text{eff}}=\lim\limits_{m\to1}\lim\limits_{n\to1}\frac{1}{1-n}\frac{\Omega_{Q'}^{2h_{g_A}} \Omega_{P'}^{2h_{g_A}} \left< \sigma_{g_A}(Q') \sigma_{g_A^{-1}}(Q) \sigma_{g_B}(P')  \sigma_{g_B^{-1}}(P) \right>_{\text{CFT}^{\bigotimes mn}}}{\Omega_{Q'}^{2nh_m} \Omega_{P'}^{2nh_m} \left< \sigma_{g_m}(Q') \sigma_{g_m}(P') \sigma_{g_m}(Q) \sigma_{g_m}(P)\right>_{\text{CFT}^{\bigotimes m}}^n}\ .
\end{equation}
As implied by the disconnected extremal surfaces shown in Fig.\ref{XXX12}, in large $c$ limit, the four-point functions in the numerator and denominator are broken into a product of a pair of two-point functions respectively, i.e.
\begin{equation}\label{515}
	S_R^{\text{eff}}=\lim\limits_{m\to1}\lim\limits_{n\to1}\frac{1}{1-n}\frac{\Omega_{Q'}^{2h_{g_A}} \Omega_{P'}^{2h_{g_A}} \left< \sigma_{g_A}(Q') \sigma_{g_A^{-1}}(Q)\right>_{mn} \left<\sigma_{g_B}(P')  \sigma_{g_B^{-1}}(P) \right>_{mn}}{\Omega_{Q'}^{2nh_m} \Omega_{P'}^{2nh_m} \left< \sigma_{g_m}(Q') \sigma_{g_m}(P')\right>_m^n \left< \sigma_{g_m}(Q) \sigma_{g_m}(P)\right>_m^n}\ .
\end{equation}
It is easy to show that the numerator and the denominator cancel out, so the effective reflected entropy simply vanishes, i.e. $S_R^{\text{eff}}=0$. Thus the final reflected entropy in this phase vanishes too.
(The correlators from the boundary description and bulk description are identified by the doubling trick. Therefore one can check that the result is consistent from either side.)
\par In summary, the reflected entropy between black hole interiors generally experiences two phases. The first phase corresponds to connected extremal surface, where the reflected entropy is given by (\ref{bulk_final_1}). When the extremal surface becomes disconnected, i.e. the island appears right in the middle of the black hole, the reflected entropy shifts to the second phase, where it simply vanishes. The time of phase transition, or Page time, is determined by the time of appearance of the island, which is given by \cite{Chu:2021gdb}
\begin{equation}\label{page_time}
	T_p=\arccosh \left( \sinh X_0 e^{\arctanh \sin\theta}\frac{2l}{\epsilon_y\cos\theta} \right)\ .
\end{equation}

\begin{figure}[htbp]
	\centering
	\includegraphics[scale=0.5]{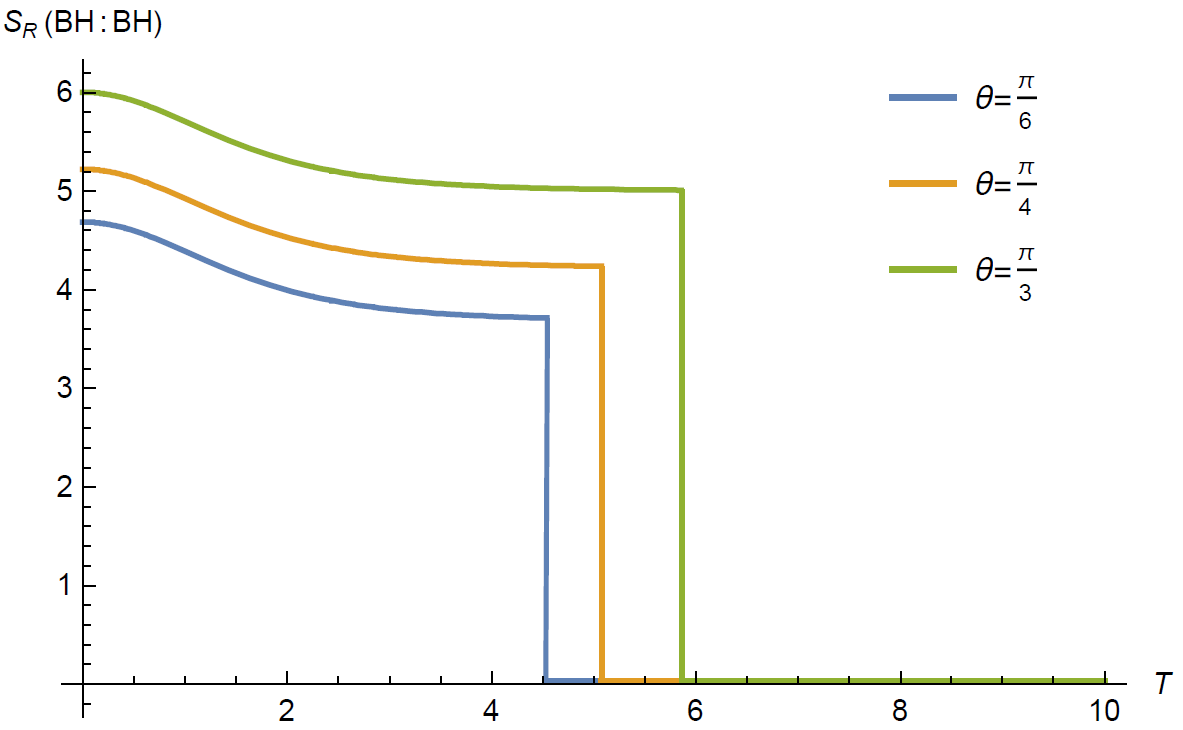}
	\caption{The reflected entropy between black hole interiors (in the unit of $\frac{c}{3}$) with respect to time $T$ for $X_0=1$ and $\theta = \frac{\pi}{6},\frac{\pi}{4},\frac{\pi}{3}$. We pick $\epsilon_y=0.1$ and $l=1$.}
	\label{XXX14}
\end{figure}
The general result under specific parameters is shown in Fig.\ref{XXX14}. One can see that the correlation between the left and right part of the black hole interior keeps decreasing over time and finally shifts to zero.

\subsubsection{Bulk description}\label{bh_bh_bulk}
In order to compute the reflected entropy between two parts of the black hole interior from the bulk side, one has to first determine the entanglement wedge of the whole black hole region and then find the minimal cross section that separates the wedge. It is easy to find the extremal surface and then the entanglement wedge of the black hole region in Euclidean geometry.

As discussed above,  The connected phase is illustrated in Fig.\ref{XXX10}, where the extremal surface is a geodesic in AdS$_3$ that connects the two endpoints on the boundary. The entanglement wedge of the black hole is bounded by the extremal surface and a space-like interval on the boundary. The next step is to find the minimal cross section of this wedge. The endpoints of the cross section are determined by extremizing over the generalized reflected entropy, i.e. the terms in bracket in (\ref{SRbulk}). This extremization procedure means that the split of the black hole on the gravitational brane is dynamical. In other words, one cannot fix the endpoint on the EOW brane by hand like in the asymptotic boundary.

\par
\begin{figure}[htbp]
  \centering
  \includegraphics[scale=0.5]{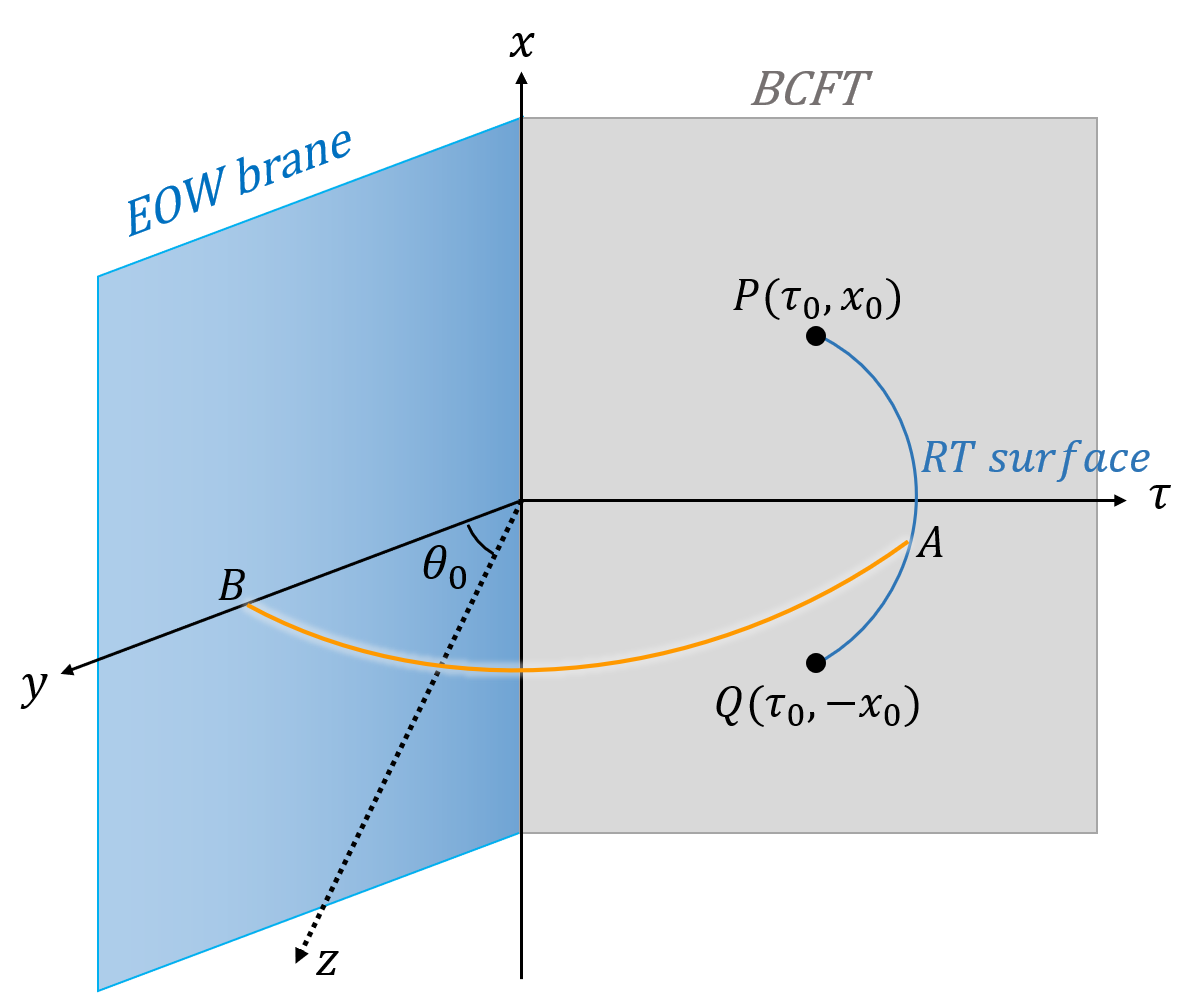}
  \caption{The connected phase in terms of $(\tau,x,z)$ coordinates. The extremal surface (RT surface) is mapped to the blue arc which lies in a plane perpendicular to the BCFT plane. The wedge cross section is represented by the orange arc, which is the minimal geodesic that connects the EOW brane and the RT surface.}
  \label{XXX11}
\end{figure}

 Now we would like to find the minimal cross section. The two endpoints of the cross section are on the EOW brane (gravitational region) and the extremal surface (Fig.\ref{XXX10}). The first term in (\ref{SRbulk}) boils down to twice the entanglement entropy of a part of the matter on the brane, which is shown to be a constant in \cite{Deng:2020ent}. Thus this effective reflected entropy reads
\begin{equation}\label{effective}
S_R^{(\text{eff})}(B_L:B_R)=\frac{c}{3}\log\frac{2l}{\epsilon_y\cos\theta}\ ,
\end{equation}
where $\epsilon_y$ is the UV cut-off on the brane. Therefore, the minimal cross section is simply the minimal geodesic connecting the EOW brane and the extremal surface. This geodesic is easier to find in the original coordinates $(\tau, x, z)$ with metrics (\ref{original_coordinates}). As shown in Fig.\ref{XXX11}, the endpoints $(\tau_0',x_0', 0)$ and $(\tau_0', -x_0', 0)$ are mapped to $(\tau_0, x_0, 0)$ and $(\tau_0, -x_0, 0)$ respectively, where
\begin{equation}
\tau_0=\frac{2(x_0'^2+\tau_0'^2-1)}{(\tau_0'+1)^2+x_0'^2}\ , \quad x_0=\frac{4x_0'}{(\tau_0'+1)^2+x_0'^2}\ .
\end{equation}

Now we have to find the minimal geodesic connecting the RT surface (the blue arc in Fig.\ref{XXX11}) and the EOW brane (the blue plane). This geodesic is part of a circle whose center lies in the $x-\tau$ plane, and it also has to be perpendicular to the plane. By observing the metrics
\begin{equation}
  ds^{2} = d\rho^{2}+l^2\cosh ^{2} \frac{\rho}{l} \cdot \frac{d\tau^{2}+dy^{2}}{y^{2}}\ ,
\end{equation}
and recalling (\ref{rhoy}) it is easy to show that the minimal condition requires the center coordinate to be $(\tau=0,x=0)$ and the endpoint on the RT surface to be right in the middle, so that $(d\tau^{2}+dy^{2})$ vanishes in the integral. Since $\rho$ satisfies $\cosh(\frac{\rho}{l}) = \frac{1}{\cos\theta}$, to make the integral of $d\rho^2$ minimize, one has to reduce the angle $\theta$ as much as possible, and it can be verified that our previous choice exactly meets this condition. In summary, the minimal arc has center coordinate $(\tau,x,z)=(0,0,0)$ and one of its endpoint A is on $(\tau,x,z)=(\tau_0, 0, x_0)$. The coordinate of the other endpoint B on the EOW brane is therefore $(-\sqrt{x_0^2+\tau_0^2}\sin\theta, 0, \sqrt{x_0^2+\tau_0^2}\cos\theta)$. With these coordinates, it is easy to obtain the length of the arc, which takes the form as
\begin{equation}\label{411}
\begin{split}
 L &= l \left [ \log \frac{\tau_0 + \sqrt{x_0^2 + \tau_0^2}}{x_0} + \log \frac{\cos \theta}{1-\sin \theta} \right ]\\
 &= l \left [ \log \frac{x_0'^2+\tau_0'^2-1+\sqrt{4 x_0'^2+(\tau_0'^2+x_0'^2-1)^2}}{2 x_0'} + \log \frac{\cos \theta}{1-\sin \theta} \right ].
\end{split}
\end{equation}
Here we have written this length formula as a function of $(\tau_0',x_0')$ in Euclidean geometry. Recall that the physical result should be read in Rindler coordinates $(X,T)$ in Lorentz geometry. Analytically continuing this result using $\tau_0'=it_0'$ together with (\ref{rindler}), and add the effective matter term (\ref{effective}), one gets the final reflected entropy between black hole interiors, which is given by
\begin{small}
	\begin{equation}\label{bulk_final_2}
		S_R^{\text{bulk}}(B_L:B_R)=\frac{c}{3}\left[ \log \frac{e^{2X_0}-1+\sqrt{4e^{2X_0}\cosh^2 T+(e^{2X_0}-1)^2}}{2e^{X_0}\cosh T} + \log \frac{\cos\theta}{1-\sin\theta}+ \log\frac{2l}{\epsilon_y\cos\theta}\right].
	\end{equation}
\end{small}
One can see that the reflected entropy above is the same as (\ref{bulk_final_1}), the result obtained in the boundary description.

\begin{figure}[htbp]
  \centering
  \includegraphics[scale=0.5]{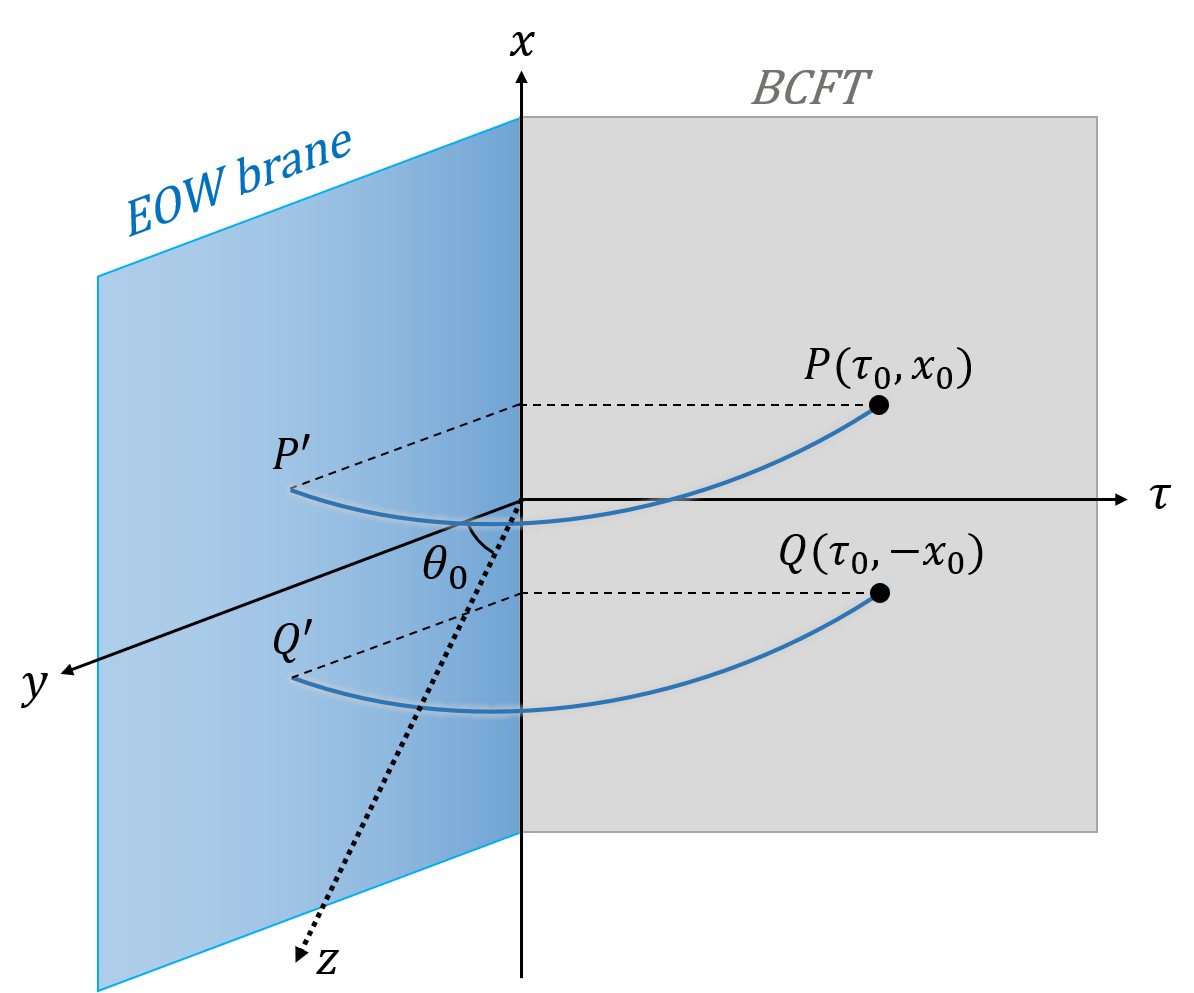}
  \caption{The disconnected phase in terms of $(\tau,x,z)$ coordinates. Extremal surfaces are represented by blue arcs.}
  \label{XXX13}
\end{figure}

The disconnected phase is shown in Fig.\ref{XXX12}. In this phase, the entanglement wedge of radiation extends to the EOW brane, which means that an island appears in the middle of the black hole region and splits the black hole into two parts. Since the entanglement wedge of the two parts are naturally separated, the area term of the entanglement wedge cross section in (\ref{SRbulk}) vanishes. The matter term in (\ref{SRbulk}) can be calculated by the correlation function of twist operators inserted at $Q'$ and $P'$ on the EOW brane,
\begin{equation}
S_R^{\text{eff}}(Q',P')=\lim\limits_{m\to1}\lim\limits_{n\to1}\frac{1}{1-n}\log \frac{\Omega_{Q'}^{2h_{g_A}} \Omega_{P'}^{2h_{g_B}} \left< \sigma_{g_A}(Q') \sigma_{g_B}(P')\right>_{\text{BCFT}^{\bigotimes mn}}}{\Omega_{Q'}^{2nh_m} \Omega_{P'}^{2nh_m} \left< \sigma_{g_m}(Q') \sigma_{g_m}(P')\right>_{\text{BCFT}^{\bigotimes m}}^n}\ . 
\end{equation}
This correlation function is easy to calculate in the $(y,x)$ coordinates, where the brane coordinates of $P'$ and $Q'$ are found to be $(y,x)=(\tau_0,x_0)$ and $(\tau_0,-x_0)$ \cite{Chu:2021gdb}, as illustrated in Fig.\ref{XXX13}. Since the functional form of a correlator of the BCFT defined in half plane has the same functional form as that of a chiral CFT on the whole plane, one can use \textit{doubling trick} to express the correlators as the following form
\begin{equation}\label{523}
S_R^{\text{eff}}(Q',P')=\lim\limits_{m\to1}\lim\limits_{n\to1}\frac{1}{1-n}\log \frac{\Omega_{Q'}^{2h_{g_A}} \Omega_{P'}^{2h_{g_B}} \left< \sigma_{g_A}(Q') \sigma_{g_A}(Q'') \sigma_{g_B}(P')  \sigma_{g_B}(P'') \right>_{\text{CFT}^{\bigotimes mn}}}{\Omega_{Q'}^{2nh_m} \Omega_{P'}^{2nh_m} \left< \sigma_{g_m}(Q') \sigma_{g_m}(Q'') \sigma_{g_m}(P') \sigma_{g_m}(P'')\right>_{\text{CFT}^{\bigotimes m}}^n}\ ,
\end{equation}
where $P''$ and $Q''$ with brane coordinates $(y,x)=(-\tau_0,x_0)$ and $(y,x)=(-\tau_0,-x_0)$ are symmetric points of $P'$ and $Q'$. It is easy to show that (\ref{523}) is essentially the same as (\ref{514}). According to the discussion in the previous section, this gives vanishing reflected entropy. Therefore, the result obtained via island formula in boundary description is justified from bulk computation.

\subsection{The reflected entropy between radiation and black hole}
In the above section we have divided the black hole into two regions denoted by $B_L$ and $B_R$ respectively. Recall that two regions are adjacent in the first phase, and then separated by an island appearing in the second phase. One can also denote the left radiation and right radiation by $R_L$ and $R_R$ respectively. The two phases are shown in Fig.\ref{XXX15} and Fig.\ref{XXX16}. In this section we would like to compute the reflected entropy between radiation and black hole in the left system. 
\subsubsection{Bulk description}
\begin{figure}[htbp]
  \centering
  \includegraphics[scale=0.5]{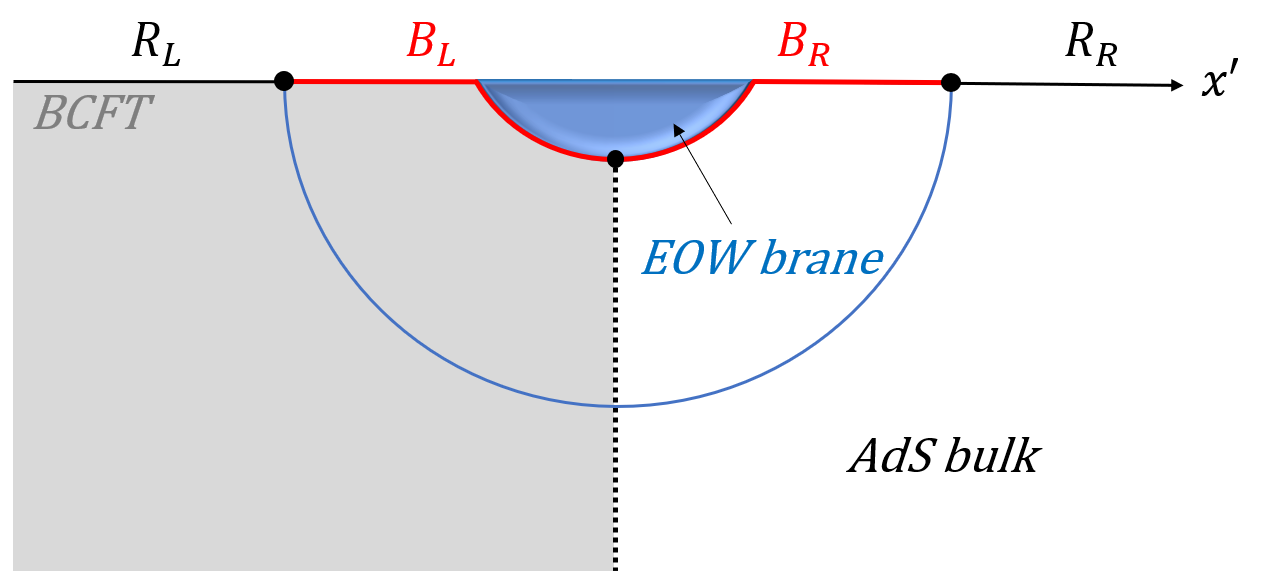}
  \caption{The connected phase viewed along $z$-axis. A time slice of black hole region is divided into $B_L$ and $B_R$. The extremal surface of the whole black hole is represented by the blue curve. The entanglement wedge of the left system $R_L\cup B_L$ is the gray shaded region, which is separated from the right part of the system by an extremal surface (dashed line).}
  \label{XXX15}
\end{figure}

To compute $S_R(R_L:B_L)$ in the bulk description, one first has to find the entanglement wedge of $R_L\cup B_L$, which is bounded by the boundary and associated extremal surface. The case for the first phase is shown in Fig.\ref{XXX15}, where the extremal surface is a geodesic that connects infinity to the intersection point of $B_L$ and $B_R$ which is found to be at $x'=0$ in the above section. The entanglement wedge is the gray shaded region. It is easy to verify that the minimal cross section between $R_L$ and $B_L$ is half of the extremal surface for $B_L\cup B_R$ by left/right $\mathbb{Z}_2$ symmetry. Thus the area term in (\ref{SRbulk}) is 
\begin{equation}
\frac{\text{Area}[\text{EW}(R_L:B_L)]}{2G_N}=\frac{\text{Area}[\text{RT}(B_L\cup B_R)]}{4G_N}\ ,
\end{equation}
where $\text{RT}(B_L\cup B_R)$ denotes the extremal surface for the entire black hole $B_L\cup B_R$. Since the matter is only located on the EOW brane from the bulk point of view, the effective reflected entropy across the wedge cross section vanishes. The final result is simply the area term
\begin{equation}\label{424}
S_R^{\text{bulk}}(R_L:B_L)=\frac{\text{Area}[\text{RT}(B_L\cup B_R)]}{4G_N}=\frac{c}{3}\log\frac{2x_0'}{\epsilon}\ ,
\end{equation}
where $\epsilon$ is the UV cut off of the asymptotic boundary.

\begin{figure}[htbp]
  \centering
  \includegraphics[scale=0.5]{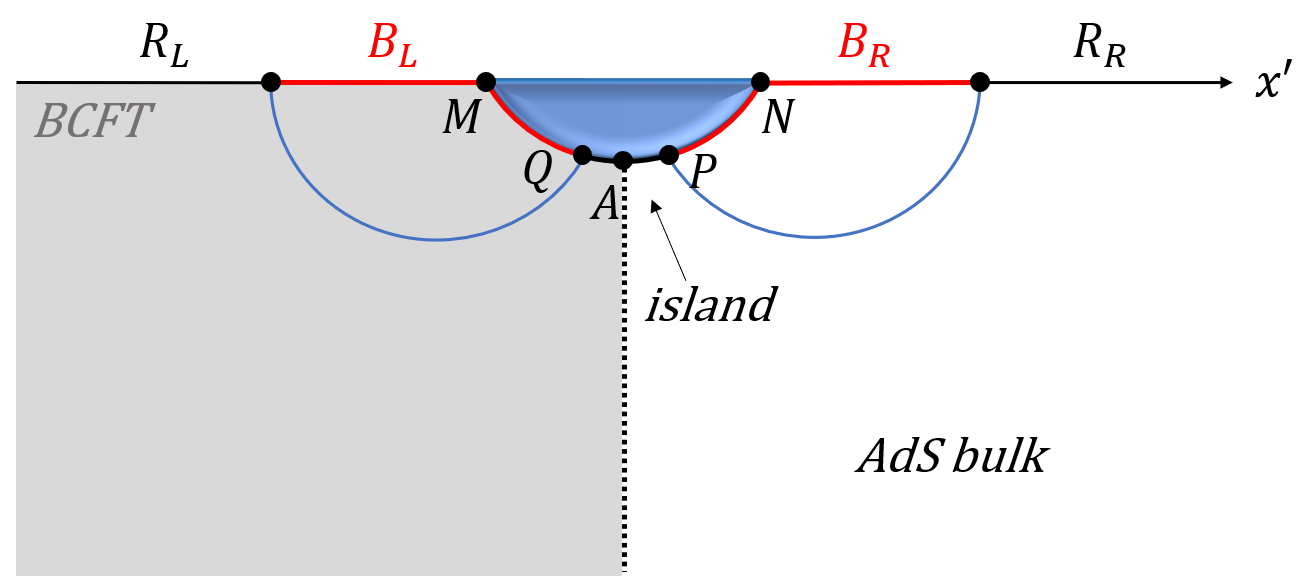}
  \caption{The disconnected phase viewed along $z$-axis. A black hole time slice has two separated regions $B_L$ and $B_R$, which are separated by an island on the EOW brane. It is obvious that the island (black curve $QP$) belongs to the entanglement wedge of radiation. The island is divided into left and right parts, which are denoted by $I_L$ and $I_R$ respectively. The entanglement wedge of the left system $R_L\cup B_L$ is again represented by the gray shaded region.}
  \label{XXX16}
\end{figure}

For the disconnected phase (Fig.\ref{XXX16}), the entanglement wedge cross section between $R_L$ and $B_L$ is naturally identified as the left part of the disconnected extremal surface for $B_L\cup B_R$ (red curve in Fig.\ref{XXX16}). Therefore, the area of the cross section is again half the area of the extremal surface for the entire black hole, which is calculated in \cite{Chu:2021gdb}. The area term for the disconnected phase is given by
\begin{equation}\label{425}
\begin{split}
\frac{\text{Area}[\text{EW}(A:B)]}{2G_N} &= \frac{\text{Area}[\text{RT}(B_L\cup B_R)]}{4G_N}\\
	&= \frac{c}{3}\left( \log\frac{x_0'^2+\tau_0'^2-1}{\epsilon}+\arctanh \sin \theta \right)\ .
\end{split}
\end{equation}

\begin{figure}[htbp]
  \centering
  \includegraphics[scale=0.5]{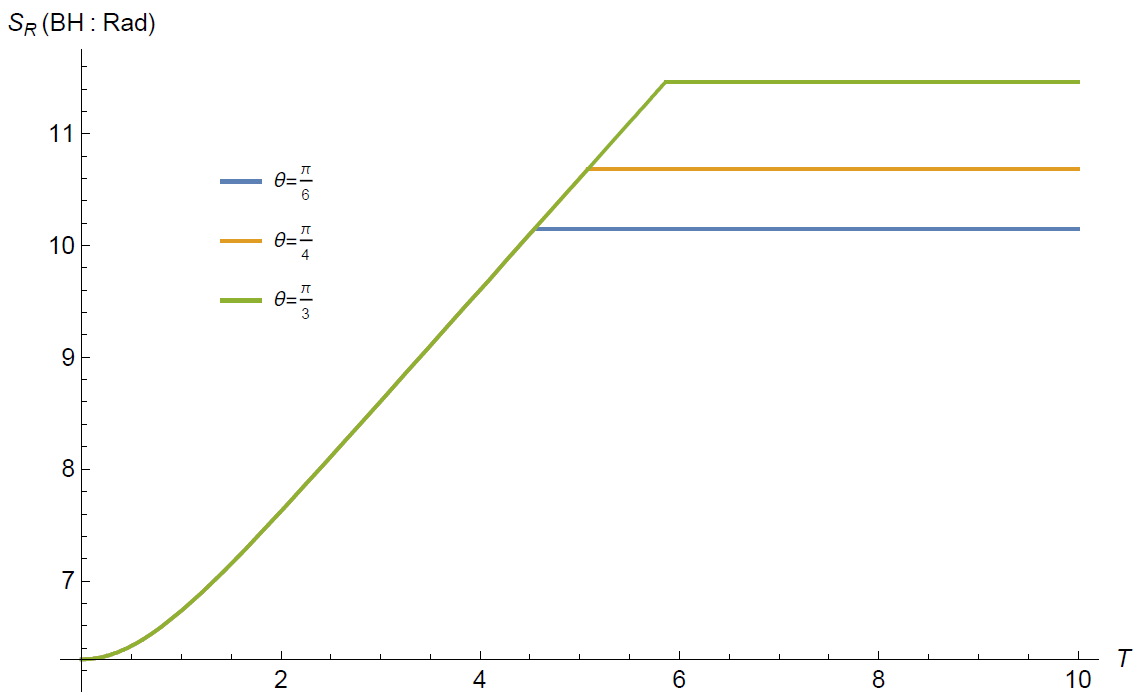}
  \caption{The reflected entropy between black hole and radiation (in the unit of $\frac{c}{3}$) with respect to time $T$ for $X_0=1$ and $\theta = \frac{\pi}{6},\frac{\pi}{4},\frac{\pi}{3}$. We pick $\epsilon=0.01$, $\epsilon_y=0.1$, and $l=1$.}
  \label{XXX17}
\end{figure}

As for the effective matter term, it computes the reflected entropy between the intervals $|QM|$ and $|QA|$ on the brane. Since a whole slice of the brane is a pure state, one can take $|AN|$ as a conanical purification of $|AM|$. Then the reflected entropy between $|QM|$ and $|QA|$ is by definition the entanglement entropy of $|QM|\cup|PN|$, which is calculated as \cite{Chu:2021gdb}
\begin{equation}
	S_R^{\text{(eff)}}(|QM|:|QA|)=S(|QM|\cup|PN|)=\frac{c}{3}\log{\frac{2l}{\epsilon_y\cos{\theta}}}.
\end{equation}
Thus, the reflected entropy in this phase is given by
\begin{equation}\label{427}
S_R^{\text{bulk}}(R_L:B_L)=\frac{c}{3}\left( \log\frac{x_0'^2+\tau_0'^2-1}{\epsilon}+\arctanh \sin \theta + \log{\frac{2l}{\epsilon_y\cos{\theta}}}\right)\ .
\end{equation}
Combining (\ref{424}) with (\ref{427}), the reflected entropy between black hole and radiation can be transformed into Rindler coordinates $(T,X)$ as
\begin{equation}
\label{tden}
\begin{split}
S_R^{\text{bulk}}(R_L:B_L)=\begin{cases}\frac{c}{3}\left(\log \frac{2\cosh T}{\epsilon}+X_0\right),\quad &T<T_P\\
\frac{c}{3}\left(\log \frac{e^{2X_0}-1}{\epsilon} + \arctanh \sin\theta + \log\frac{2l}{\epsilon_y\cos\theta}\right)\ , \quad &T>T_P,
\end{cases}
\end{split}
\end{equation}
where $T_P$ is the Page time given by (\ref{page_time}). The time dependent reflected entropy is plotted in Fig.\ref{XXX17}. It is worth noting that it follows the same Page curve of the entanglement entropy of the whole black hole or radiation.

\subsubsection{Boundary description}
When computing the reflected entropy between left radiation and left black hole, the boundary island formula (\ref{SRbdy}) becomes
\begin{equation}
S_R^{\text{bdy}}(R_L:B_L)=\text{min}\text{Ext}_{\Gamma}\left\{S_R^{(\text{eff})}(R_L\cup I_L:B_L)+\frac{\text{Area}[\Gamma=\partial I_L \cap \partial B_L]}{2G_N}\right\}\ .
\end{equation}
When $I_L=\emptyset$, this corresponds to the first case, where the area term vanishes and the first term boils down to $S_R^{(\text{eff})}(R_L:B_L)$, which is just the entanglement entropy of $B_L\cup B_R$ via canonical purification. This can be calculated by a two-point correlator of twist operators inserting at $(\tau_0',x_0')$ and $(\tau_0', -x_0')$, and the result is
\begin{equation}
S_R^{\text{bdy}}(R_L:B_L)=\frac{c}{3}\log\frac{2x_0'}{\epsilon}\ ,
\end{equation}
which is the same as (\ref{424}). When $I_L$ is not empty set, according to (\ref{art}), the area term is given by
\begin{equation}
	\frac{\text{Area}[\Gamma]}{2G_N}=\frac{c}{3}\arctanh \sin\theta\ .
\end{equation}
The matter term $S_R^{(\text{eff})}(R_L\cup I_L:B_L)$ can also be canonically purified as the entanglement entropy of $B_L\cup B_R$, which can be found in \cite{Chu:2021gdb}. The result is
\begin{equation}
S_R^{(\text{eff})}(R_L\cup I_L:B_L)=\frac{c}{3}\log\frac{2l(x_0'^2+\tau_0'^2-1)}{\cos\theta\epsilon\epsilon_y}\ .
\end{equation}
In summary, the reflected entropy for this phase reads
\begin{equation}
S_R^{\text{bdy}}(R_L:B_L)=\frac{c}{3}\left( \arctanh \sin\theta + \log \frac{2l(x_0'^2+\tau_0'^2-1)}{\cos\theta\epsilon\epsilon_y}  \right)\ ,
\end{equation}
which coincides with (\ref{427}) exactly.
\subsection{The reflected entropy between radiation and radiation}
In this section we study the reflected entropy between two intervals of radiation, i.e. $R_L$ and $R_R$, as shown in Fig.\ref{XXX20} and Fig.\ref{XXX19}. For the interval $R_R$ in the right Rindler patch, the endpoints are specified as $(T,X_0)$ and $(T,X_1)$ in $(T,X)$ coordinates. Again, we will do the computation in Euclidean geometry (Fig.\ref{XXX20} and Fig.\ref{XXX19}), where the four endpoints of the two intervals are mapped to $(\tau_0',x_0')$, $(\tau_0',-x_0')$, $(\tau_1',x_1')$ and $(\tau_1',-x_1')$ via the following transformations
\begin{equation}
x_0'=e^{X_0}\cosh T\ , \quad \tau_0'=ie^{X_0}\sinh T\ , \quad x_1'=e^{X_1}\cosh T\ , \quad \tau_1'=ie^{X_1}\sinh T\ .
\end{equation}
\subsubsection{Boundary description}
\begin{figure}[htbp]
	\centering
	\includegraphics[scale=0.5]{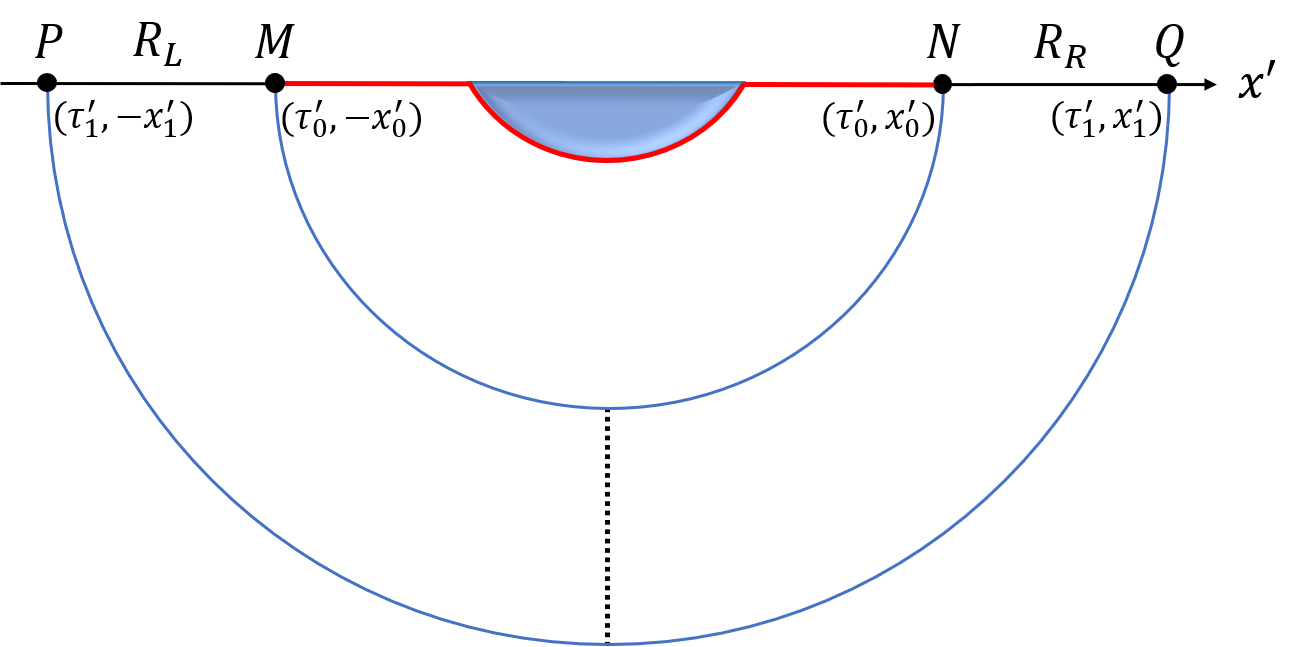}
	\caption{The connected phase viewed along $z$-axis. In this phase, the island is empty set. The reflected entropy between $R_L$ and $R_R$ is equal to the area of the entanglement wedge cross section, which is represented by the dashed line.}
	\label{XXX20}
\end{figure}

The boundary island formula for the reflected entropy between two intervals $R_L$ and $R_R$ is given by
\begin{equation}
	S_R^{\text{bdy}}(R_L:R_R)=\text{min }\text{Ext}_{\Gamma}\left\{S_R^{(\text{eff})}(R_L\cup I_L:R_R\cup I_R)+\frac{\text{Area}[\Gamma=\partial I_L \cap \partial I_R]}{2G_N}\right\}\ .
\end{equation}
For the connected phase shown in Fig.\ref{XXX20}, the island is empty set, so one only needs to compute the reflected entropy between two intervals on a flat CFT, whose result can be found in \cite{Dutta:2019gen}
\begin{small}
	\begin{equation}\label{REofRRbdy}
		\begin{split}
			S_R^{\text{bdy}}(R_L:R_R)=S_R^{(\text{eff})}(R_L:R_R)=\frac{2c}{3}\log \frac{1+\sqrt{1-\eta}}{\sqrt{\eta}}\ , \\\eta=\frac{|MN||PQ|}{|PN||MQ|}=\frac{4e^{X_0+X_1}}{(e^{X_0}+e^{X_1})^2-(e^{X_0}-e^{X_1})^2\tanh^2T},
		\end{split}
	\end{equation}
\end{small}
which is a monotonically decreasing function of physical time $T$.

\begin{figure}[htbp]
	\centering
	\includegraphics[scale=0.5]{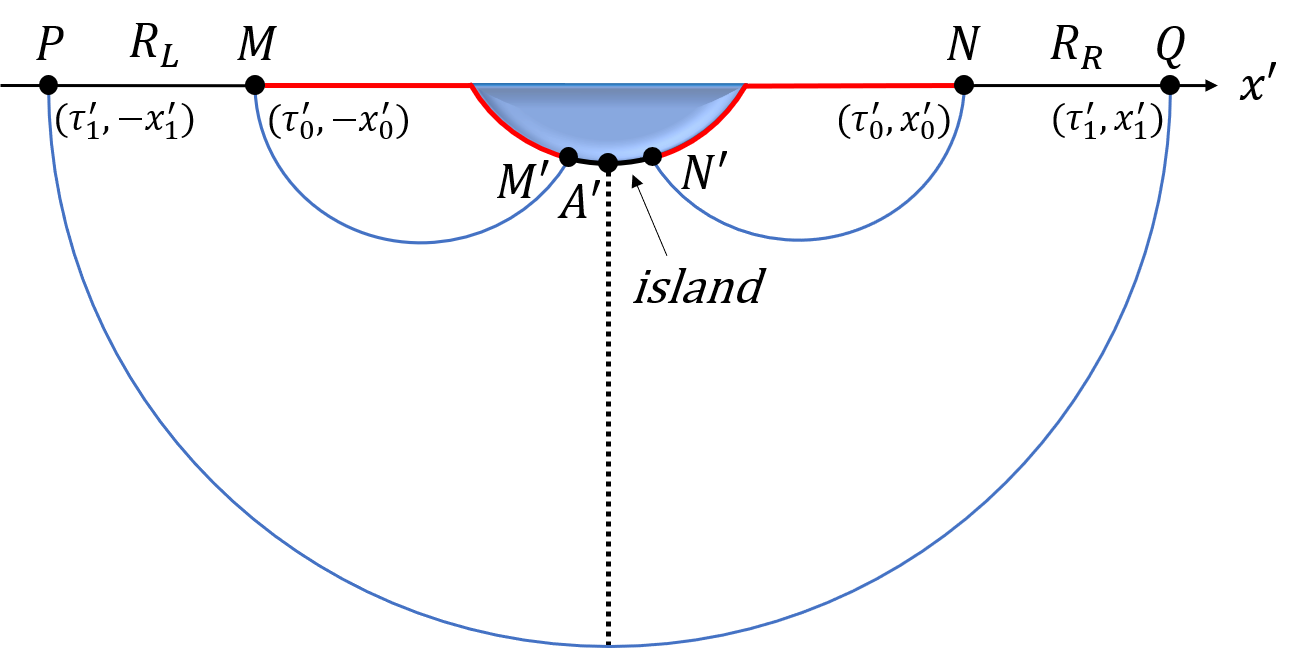}
	\caption{The disconnected phase viewed along $z$-axis. In this phase, the reflected entropy between $R_L$ and $R_R$ should include the contribution from the island, the interval on the brane bounded by $M'$ and $N'$. From the bulk perspective, the reflected entropy is the area of the wedge cross section (dashe line) plus the effective reflected entropy between $|M'A'|$ and $|N'A'|$.}
	\label{XXX19}
\end{figure}

For the disconnected phase illustrated in Fig.\ref{XXX19}, the area term is $\frac{c}{3}\arctanh \sin\theta$. The effective entropy term is given by
\begin{small}
	\begin{equation}\label{5391}
		\begin{split}
			S_R^{(\text{eff})}&=\lim\limits_{m\to1}\lim\limits_{n\to1}\frac{1}{1-n}\log \frac{\Omega_{M'}^{2h_A} \Omega_{A}^{2h_{AB}} \Omega_{N'}^{2h_B} \left< \sigma_A(P) \sigma_{A^{-1}}(M)\sigma_A(M')\sigma_{A^{-1}B}(A')\sigma_{B^{-1}}(N')\sigma_{B}(N)\sigma_{B^{-1}}(Q) \right>_{\text{CFT}^{\bigotimes mn}}  }{\Omega_{M'}^{2nh_m} \Omega_{N'}^{2nh_m} \left< \sigma_m(P) \sigma_m(M)\sigma_m(M')\sigma_m(N')\sigma_m(N)\sigma_m(Q) \right>_{\text{CFT}^{\bigotimes m}}^n }\\
			&=\lim\limits_{m\to1}\lim\limits_{n\to1}\frac{1}{1-n}\log \frac{\Omega_{A}^{2h_{AB}} \left< \sigma_{A^{-1}}(M) \sigma_A(M')\right>_{mn} \left< \sigma_{B^{-1}}(N')\sigma_{B}(N) \right>_{mn} \left< \sigma_A(P) \sigma_{B^{-1}}(Q)\sigma_{A^{-1}B}(A') \right>_{mn} }{\left< \sigma_m(M)\sigma_m(M') \right>_m^n \left<\sigma_m(N')\sigma_m(N) \right>_m^n \left<\sigma_m(P)\sigma_m(Q) \right>_m^n }\\
			&=\lim\limits_{m\to1}\lim\limits_{n\to1}\frac{1}{1-n}\log \frac{\Omega_{A}^{2h_{AB}} \left< \sigma_A(P) \sigma_{B^{-1}}(Q)\sigma_{A^{-1}B}(A') \right>_{mn} }{\left<\sigma_m(P)\sigma_m(Q) \right>_m^n }\ ,
		\end{split}
	\end{equation}
\end{small}
where in the second line we have let the correlators factorize into their respective contractions\footnote{The contractions can be justified by the holographic extremal surfaces shown in Fig.\ref{XXX19}.} assuming large $c$ limit \cite{Hartman:2013mia}. The third line turns out to take the same form as (\ref{SReffbdy_bh}) with $x_0'$ replaced by $x_1'$, which essentially computes the reflected entropy between two adjacent intervals . Thus, one can directly modify (\ref{bulk_final_1}) to obtain the result
\begin{small}
	\begin{equation}\label{538}
		S_R^{\text{bdy}}(R_L:R_R)=\frac{c}{3}\left[ \log \frac{e^{2X_1}-1+\sqrt{4e^{2X_1}\cosh^2 T+(e^{2X_1}-1)^2}}{2e^{X_1}\cosh T} + \log \frac{\cos\theta}{1-\sin\theta} + \log\frac{2l}{\epsilon_y\cos\theta}\right].
	\end{equation}
\end{small}

\par
\begin{figure}[htbp]
	\centering
	\includegraphics[scale=0.5]{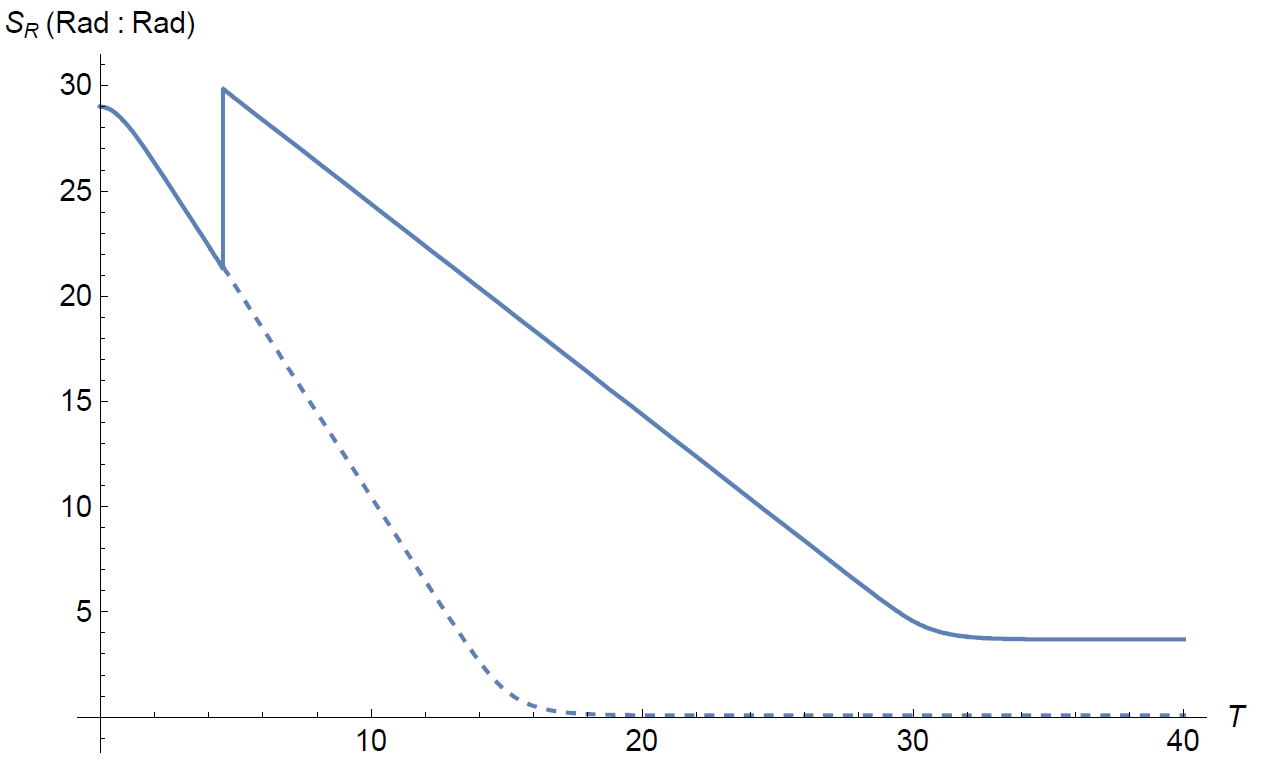}
	\caption{The reflected entropy between radiation and radiation (in the unit of $\frac{c}{3}$) with respect to time $T$ for $X_0=1$. We pick $X_1=30$, $\theta = \frac{\pi}{6}$, $\epsilon_y=0.1$, and $l=1$. The solid line represents the time evolution of the final reflected entropy while the dotted line represents the result for the first phase. Note that the R-R reflected entropy jumps at Page time with a gap given by (\ref{gap}) in the limit $X_1 \rightarrow \infty$.}
	\label{XXX21}
\end{figure}

 The R-R reflected entropy is plotted in Fig.\ref{XXX21}. Note that generally the entropy decreases with time in both phases, but it jumps at Page time when the phase transition occurs. We are especially interested in the case where both the left and right radiation extend to spacial infinity, i.e. $X_1\rightarrow \infty$. In this limit, the asymptotic behavior of the reflected entropy is given by
\begin{equation}
	\begin{split}
		S_R^{\text{bdy}}(R_L:R_R)=\begin{cases}\frac{c}{3}\left(  X_1-X_0-2\log\cosh T     \right)\ ,\quad &T<T_P\\
			\frac{c}{3}\left(X_1 - \log\cosh T + \log \frac{\cos\theta}{1-\sin\theta} + \log \frac{2l}{\epsilon_y\cos\theta} \right)\ ,\quad &T>T_P\ .
		\end{cases}
	\end{split}
\end{equation}
One can see that the gap at $T_P$ in this limit is
\begin{equation}\label{gap}
	S_{\text{gap}}=\frac{c}{3}\left( 2\log \frac{2l}{\epsilon_y\cos\theta} + 2\log \frac{\cos\theta}{1-\sin\theta} + \log \frac{e^{2X_0}-1}{2} \right)\ .
\end{equation}

\subsubsection{Bulk description}
For the connected phase (Fig.\ref{XXX20}), the entanglement wedge of $R_L\cup R_R$ is bounded by two extremal surfaces. There is no matter contribution within the wedge, so one only needs to compute the area of the wedge cross section that separates $R_L$ and $R_R$, which is the minimal geodesic connects the two extremal surfaces. 

\par
\begin{figure}[htbp]
	\centering
	\includegraphics[scale=0.5]{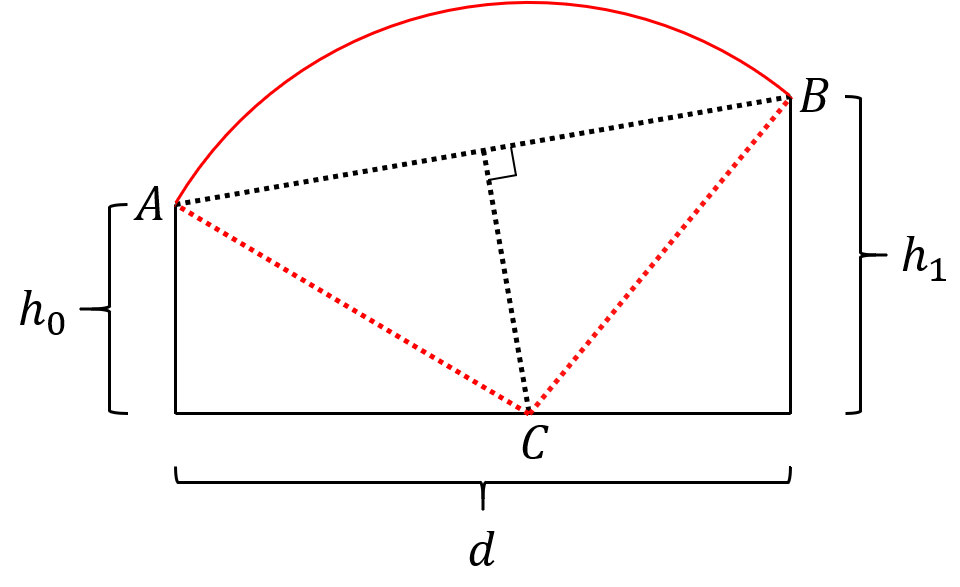}
	\caption{A slice perpendicular to the $\tau'Ox'$ plane. The solid red line is the geodesic, $C$ is the center of geodesic, $A$ and $B$ are the ending points on the RT surfaces, $h_0$ and $h_1$ are the values of the $z$ coordinate of $A$ and $B$, $d$ is the distance of the projection of $A$ and $B$ on $\tau'Ox'$ plane. }
	\label{SLICE2}
\end{figure}

 The computation in the coordinates $(\tau',x',z')$ is as follows. First we calculate a general expression of the length of the geodesic shown in Fig.\ref{SLICE2},
which is given by
\begin{equation}
\begin{split}
L(h_0,h_1,d) 
&= \log\frac{h_0^2+h_1^2+d^2+\sqrt{(h_0^2+h_1^2+d^2)^2-4h_0^2h_1^2}}{2h_0h_1}\ , \\
&= \arccosh\frac{h_0^2+h_1^2+d^2}{2h_0h_1}\ .
\end{split}
\end{equation}
The coordinate of A is $(\tau',x',z')=(\tau'_0,a,\sqrt{{x'_0}^2 - a^2})$ and for B we have $(\tau',x',z')=(\tau'_1,b,\sqrt{{x'_1}^2 - b^2})$. Employing $h_0 = \sqrt{{x'_0}^2 - a^2}$, $h_1 = \sqrt{{x'_1}^2 - b^2}$ and $d = \sqrt{(a-b)^2+(\tau'_1-\tau'_0)^2}$, we get
\begin{equation}
	L(a,b) = \arccosh\frac{x_0'^2+x_1'^2-2ab+(\tau'_1-\tau'_0)^2}{2\sqrt{{x'_0}^2 - a^2}\sqrt{{x'_1}^2 - b^2}}\ .
\end{equation}
It can be checked that $L(a,b)$ reaches its minimum when $a=b=0$, so the minimal geodesic is
\begin{equation}
	L_{\text{min}} = \arccosh\frac{x_0'^2+x_1'^2+(\tau'_1-\tau'_0)^2}{2x'_0x'_1}\ .
\end{equation}
Thus, the reflected entropy written in Rindler coordinates is given by
\begin{equation}\label{REofRRBulk}
	S_R^{\text{bulk}}(R_L:R_R)=\frac{L_{\text{min}}}{2G_N}=\frac{c}{3} \arccosh \frac{e^{X_0 - X_1} + e^{X_1 - X_0} + 2\sinh^2T}{2\cosh^2T} \ .
\end{equation}
One can check that (\ref{REofRRBulk}) agrees with (\ref{REofRRbdy}) exactly. 

\par For the disconnected phase (Fig.\ref{XXX19}), the entanglement wedge cross section is a geodesic connecting the larger extremal surface and the EOW brane. The intersection point of the cross section and the brane, which is denoted by $A'$, is determined by minimizing the general reflected entropy. Here one needs to include the contribution from effective matter on the brane, which can be calculated as follows
\begin{small}
	\begin{equation}
		\begin{split}
			S_R^{(\text{eff})}(|A'M'|:|A'N'|)&=\lim\limits_{m\to1}\lim\limits_{n\to1}\frac{1}{1-n}\log \frac{ \Omega_{M'}^{2h_A} \Omega_{A'}^{2h_{AB}} \Omega_{N'}^{2h_B} \left< \sigma_A(M') \sigma_{AB}(A') \sigma_B(N') \right>_{\text{BCFT}^{\bigotimes mn} }}{ \Omega_{M'}^{2nh_m} \Omega_{N'}^{2nh_m} \left< \sigma_m(M') \sigma_m(N') \right>_{\text{BCFT}^{\bigotimes m} }^n}\\
			&=\lim\limits_{m\to1}\lim\limits_{n\to1}\frac{1}{1-n}\log \frac{ \Omega_{A'}^{2h_{AB}}  \left< \sigma_A(M') \sigma_A(M) \sigma_{AB}(A') \sigma_{AB}(A) \sigma_B(N') \sigma_B(N) \right>_{\text{CFT}^{\bigotimes mn} }}{  \left< \sigma_m(M') \sigma_m(M) \sigma_m(N') \sigma_m(N) \right>_{\text{CFT}^{\bigotimes m} }^n}\\
			&=\lim\limits_{m\to1}\lim\limits_{n\to1}\frac{1}{1-n}\log \frac{ \Omega_{A'}^{2h_{AB}} \left< \sigma_A(M') \sigma_A(M)  \right>_{mn} \left< \sigma_B(N') \sigma_B(N) \right>_{mn} \left< \sigma_{AB}(A') \sigma_{AB}(A) \right>_{mn} }{   \left< \sigma_m(M') \sigma_m(M) \right>_m^n \left< \sigma_B(N') \sigma_B(N) \right>_m^n }\\
			&=\lim\limits_{m\to1}\lim\limits_{n\to1}\frac{1}{1-n}\log(\Omega_{A'}^{2h_{AB}} \left< \sigma_{AB}(A') \sigma_{AB}(A) \right>_{\text{CFT}^{\bigotimes mn}} )\\
			&=\lim\limits_{m\to1}\lim\limits_{n\to1}\frac{1}{1-n} \log( \Omega_{A'}^{2h_{AB}} \left< \sigma_{AB}(A') \right>_{\text{BCFT}^{\bigotimes mn}})\ .
		\end{split}
	\end{equation}
\end{small}
In the second line, the doubling trick is employed and the warp factors cancel out. In the third line the correlators are factorized into contractions following the same channel as that in (\ref{5391}). In the last line we recover a one-point correlator of the BCFT on the EOW brane by reversing the doubling trick. Since $h_{AB}=\frac{c}{12}(n-\frac{1}{n})$ equals twice the conformal dimension of the twist operator for entanglement entropy, the last line is essentially twice the entanglement entropy of half of a slice $[A',\infty]$ in the BCFT, which can be found in \cite{Deng:2020ent}. The result reads
\begin{equation}
	S_R^{(\text{eff})}(|A'M'|:|A'N'|)=2S([A',\infty])=\frac{c}{3}\log\frac{2l}{\epsilon_y\cos\theta}\ .
\end{equation}
Note that it is a constant, thus minimizing the generalized reflected entropy is equivalent to minimizing the area of the wedge cross section. Therefore, the cross section is simply the minimal geodesic, whose length can be shown to take the same form as (\ref{411}) with $x_0'$ and $\tau_0'$ replaced by $x_1'$ and $\tau_1'$ respectively. To summarize, the final reflected entropy turns out to be
\begin{small}
	\begin{equation}\label{437}
		S_R^{\text{bulk}}(R_L:R_R)=\frac{c}{3}\left[ \log \frac{e^{2X_1}-1+\sqrt{4e^{2X_1}\cosh^2 T+(e^{2X_1}-1)^2}}{2e^{X_1}\cosh T} + \log \frac{\cos\theta}{1-\sin\theta}+ \log\frac{2l}{\epsilon_y\cos\theta}\right],    
	\end{equation}
\end{small}
which is the same as (\ref{538}).
\section{Conclusion and discussion}

In this paper we study the holographic dual of reflected entropy in models with defects. We particularly focus on the AdS/BCFT model which includes a brane defect in the bulk. Including the contribution from the defect theory on the brane, we propose the defect extremal surface formula for reflected entropy, which we called defect extremal cross section. On the other hand, this model is tightly related to a lower dimensional gravity system glued to a quantum bath. In fact there is a concrete procedure, including both Randall-Sundrum and Maldacena duality, to give a lower dimensional effective description for the same system. We demonstrate the equivalence between defect extremal cross section formula and boundary island formula for reflected entropy in AdS$_3$/BCFT$_2$. Extending the study to time dependent case, we also find that the bulk formula and boundary island formula give precisely the same results, including the black hole-black hole reflected entropy, black hole-radiation reflected entropy and radiation-radiation reflected entropy.

There are a few future questions listed in order: First, generalize our study to higher dimensions. In higher dimensions, defect extremal surface may not give precisely the same results as island formula~\cite{Chu:2021gdb}. It would be interesting to understand the discrepancy by looking into the reflected entropy. Second, generalize our study to AdS with other defects. One interesting example would be Wilson loop. We expect that our defect extremal surface formula also works for other defects. Last, generalize our study to multipartite correlations. Following the multipartite generalizations of reflected entropy \cite{Chu:2019etd,Umemoto:2018jpc}, we can try to find the island formula for generalized reflected entropy. This would be particularly useful in understanding the rich correlations in Hawking radiations, and therefore the nature of the island itself.
\section*{Acknowledgements}
We are grateful for useful discussions with Yilu Shao, Jinwei Chu and other group members in Fudan University. This work is supported by NSFC grant 11905033. YZ is also supported by NSFC 12047502,11947301 through Peng Huanwu Center for Fundamental Theory.
\appendix
\section{The length formula of $L(a)$ and its extremal solution}\label{appendix1}
\par
\begin{figure}[htbp]
  \centering
  \includegraphics[scale=0.45]{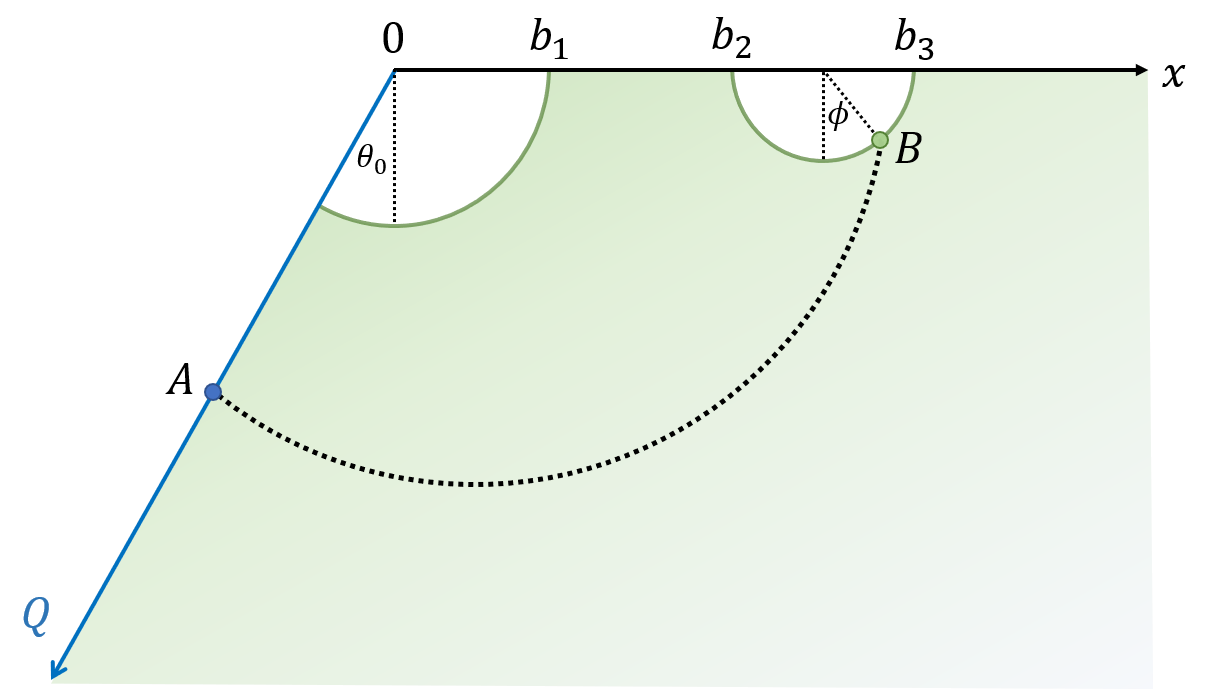}
  \caption{Here we assume $\phi$ is positive in the case shown and negative if B is on the left of the vertical dotted line, and we assume $\theta_0$ is a positive constant.}
  \label{XXX22}
\end{figure}

In this appendix we compute the length $L(a)$ of the geodesic connecting the RT surface and a fixed point A on the brane (Fig.\ref{XXX4}). We redrawn Fig.\ref{XXX4} in Fig.\ref{XXX22} to make the following calculation more clear. 

The coordinate of a point B on the RT surface is
\begin{gather}
z_B = r \cos\phi\ ,\\
x_B = R + r \sin\phi\ ,
\end{gather}
where $r=\frac{1}{2}(b_3-b_2)$ and $R=\frac{1}{2}(b_2+b_3)$. And the coordinate of A reads
\begin{gather}
z_A = a \cos\theta_0\ ,\\
x_A = -a \sin\theta_0\ .
\end{gather}
This geodesic is a circle whose center is on the horizontal axis, the coordinate of which is given by
\begin{gather}
x_{O'}=\frac{-a^2+r^2+R^2+2rR\sin\phi}{2(R + a\sin\theta_0 + r\sin{\phi})}\ ,\\
z_{O'}=0\ .
\end{gather}
The radius of the geodesic is given by
\begin{equation}
R_g=\sqrt{a^2\cos^2\theta_0+(a\sin\theta_0+\frac{-a^2+r^2+R^2+2rR\sin\phi}{2(R+a\sin\theta_0+r\sin\phi)})^2}\ .
\end{equation}
Using the coordinates above and the metric of $AdS_2$ one can get the length formula of the geodesic as a function of the coordinate of A and B, which reads
\begin{equation}\label{Laphi}
L(a,\phi)=l\arctanh{\frac{\sin\theta_0}{\sqrt{1 - f(\phi)^2 \cos^2\theta}}}+l\arctanh{\frac{\sin\theta'}{\sqrt{1 - f(\phi)^2 \cos^2\theta'}}}\ ,
\end{equation}
where $f(\phi)={x_{O'}}/{R_g}$, $\theta'=\arctan(x_B/z_B)$. It can be checked that if we fix $a$, $L(a,\phi)$ reaches its minimal value when the geodesic is perpendicular to the RT surface at the intersecting point B, namely
\begin{equation}\label{minicd}
  r\tan(\frac{\pi}{2}+\phi)=R_g\ .
\end{equation}
Substitute (\ref{minicd}) into (\ref{Laphi}), we get the length formula of the geodesic as a function of the coordinate of A, which reads
\begin{gather}
\begin{split}
L(a)=&l\arctanh{\frac{\sin\theta_0}{\sqrt{1-\frac{(a^2+r^2-R^2)^2\cos^2\theta_0}{(a^2+r^2+R^2+2 a r \sin^2\theta_0)^2-4 r^2(R+a\sin\theta_0)^2}}}} \\
&+l\arctanh{\frac{\sin\theta'}{\sqrt{1-\frac{(a^2+r^2-R^2)^2\cos^2\theta'}{(a^2+r^2+R^2+2 a r \sin^2\theta_0)^2-4 r^2(R+a\sin\theta_0)^2}}}}\ ,
\end{split}\\
\theta'=\arctan{\frac{\frac{R}{r}(a^2+r^2+R^2+2aR\sin\theta_0)-2r(R+a\sin\theta_0)}{\sqrt{(a^2+r^2+R^2+2aR\sin\theta_0)^2-4r^2(R+a\sin\theta_0)^2}}}\ .
\end{gather}
Extremizing $L(a)$ over $a$ gives the extremal solution
\begin{equation}\label{miniofg}
L_{min}=L(a_0)=l\arctanh\sin\theta_0+l\arctanh\frac{2\sqrt{b_2b_3}}{b_2+b_3}\ , \quad a_0=\sqrt{b_2b_3}\ .
\end{equation}
Actually one could find this minimal solution by a direct analysis of the metric
\begin{equation}
ds^2 = d\rho^2 + \frac{l^2}{\cos^2\theta}\frac{dy^2}{y^2}\ ,
\end{equation}
where $\rho$ satisfies $\cosh(\frac{\rho}{l}) = \frac{1}{\cos\theta}$. The length of the geodesic is
\begin{equation}
    L = \int_{-{\theta}_{0}}^{\frac{\pi}{2}-{\theta}_{1}}ds = \int_{-{\theta}_{0}}^{\frac{\pi}{2}-{\theta}_{1}}\sqrt[]{d\rho^2+{\frac{l^2}{\cos^2\theta}}\frac{dy^2}{y^2}}\ ,
\end{equation}
Where $\rho$ satisfies $\cosh(\frac{\rho}{l}) = \frac{1}{\cos\theta}$. In order to find the minimal value, one must let $y$ be a constant, which means the center of this geodesic is located at the origin of this coordinate system. Thus we have
\begin{equation}
    L \ge \int_{-{\theta}_{0}}^{\frac{\pi}{2}-{\theta}_{1}} |d\rho| = L'\ ,
\end{equation}
where $\theta_1$ is the angle between OB and $x$-axis, i.e. $\theta_1=\arctan z_B/x_B$. When $\theta$ changes from $-\theta_0$ to $0$, the value of $\rho$ decreases, so $\int_{-{\theta}_{0}}^{0} |d\rho| = \rho(-\theta_0)-\rho(0)$. Similarly we get $\int_{0}^{\frac{\pi}{2}-\theta_1} |d\rho| = \rho({\frac{\pi}{2}-\theta_1})-\rho(0)$. So we have
\begin{equation}
    L'= \rho({\frac{\pi}{2}-\theta_1})-\rho(0)+\rho(-\theta_0)-\rho(0) = l\arccosh\frac{1}{\cos\theta_0} + l\arccosh\frac{1}{\cos\theta_1}\ .
\end{equation}
The value of $L'$ depends on $\theta_1$ only. When $\theta_1$ takes the maximum value $\arcsin{\frac{R-r}{R+r}}$, $L'$ takes the minimum value
\begin{equation}\label{Lmin}
L(a_0) = l\arccosh\frac{1}{\cos\theta_0} + l\arccosh\frac{b_3+b_2}{b_3-b_2}\ ,
\end{equation}
which agrees with (\ref{miniofg}) exactly. It is easy to find that the minimal value of $\theta_1$ corresponds to the case that |OB| is tangent to the RT surface (Fig.\ref{XXX4}). Therefore the coordinate of A on the brane is simply the length of |OB|,
\begin{equation}
a_0=\sqrt{b_2 b_3}\ .
\end{equation}

\end{document}